\documentclass[fleqn,usenatbib]{mnras}

\usepackage{newtxtext,newtxmath}

\usepackage[T1]{fontenc}

\DeclareRobustCommand{\VAN}[3]{#2}
\let\VANthebibliography\thebibliography
\def\thebibliography{\DeclareRobustCommand{\VAN}[3]{##3}\VANthebibliography}

\usepackage{graphicx}	
\usepackage{amsmath}	
\usepackage{CJK}
\usepackage[whole]{bxcjkjatype}
\usepackage{CJKutf8}
\usepackage{lineno}

\newcommand{\NaiD}{Na~{\sc i}~{\sc d}}

\newcommand{\Feii} {\ion{Fe}{ii}}
\newcommand{\FeiiF} {[\ion{Fe}{ii}]}

\newcommand{\CaiiF} {[\ion{Ca}{ii}]}
\newcommand{\Ci} {\ion{C}{i}}
\newcommand{\Cii} {\ion{C}{ii}}

\newcommand{\Hei} {\ion{He}{i}}

\newcommand{\Nii} {[\ion{N}{ii}]}

\newcommand{\SiII} {\ion{Si}{ii}}

\newcommand{\Mgii} {\ion{Mg}{ii}}

\newcommand{\CoiiiF}{[\ion{Co}{iii}]}

\newcommand{\NiIIF}{[\ion{Ni}{ii}]}
\newcommand{\Feiii} {\ion{Fe}{iii}}

\newcommand{\kms}{km s$^{-1}$}
\newcommand{\nick}{$^{56}$Ni}
\newcommand{\msun}{\mbox{M$_{\odot}$}}
\newcommand{\zsun}{\mbox{Z$_{\odot}$}}

\title[SN 2020udy: deflagration of a CO white dwarf]{SN 2020udy: a SN Iax with strict limits on interaction consistent with a helium-star companion}

\author[K. Maguire et al.]{Kate Maguire$^{1}$\thanks{E-mail:kate.maguire@tcd.ie }, 
Mark R. Magee$^{2}$,
Giorgos Leloudas$^{3}$,
Adam A. Miller$^{4,5}$,
Georgios Dimitriadis$^{1}$,
\newauthor
Miika Pursiainen$^{3}$,
Mattia Bulla$^{6,7,8}$,
Kishalay De$^{9}$,
Avishay Gal-Yam$^{10}$,
Daniel A. Perley$^{11}$, 
\newauthor
Christoffer Fremling$^{12,13}$,
Viraj R. Karambelkar$^{13}$,
Jakob Nordin$^{14}$,
Simeon Reusch$^{14,15}$,
Steve Schulze$^{16}$,
\newauthor
Jesper Sollerman$^{17}$,
Giacomo Terreran$^{18}$,
Yi Yang\begin{CJK*}{UTF8}{gbsn} (杨轶)\end{CJK*}$^{19}$,
Eric C. Bellm$^{20}$,
Steven L. Groom$^{21}$,
\newauthor
Mansi M. Kasliwal$^{13}$,
Shrinivas R. Kulkarni$^{13}$,
Leander Lacroix$^{22,16}$,
Frank J. Masci$^{21}$,
Josiah N. Purdum$^{12}$,
\newauthor
Yashvi Sharma$^{13}$,
Roger Smith$^{12}$ \\
$^{1}$School of Physics, Trinity College Dublin, The University of Dublin, Dublin 2, Ireland\\
$^{2}$Department of Physics, University of Warwick, Gibbet Hill Road, Coventry CV4 7AL, UK \\
$^{3}$DTU Space, National Space Institute, Technical University of Denmark, Elektrovej 327, DK-2800 Kgs. Lyngby, 
Denmark \\
$^{4}$Department of Physics and Astronomy, Northwestern University, 2145 Sheridan Rd, Evanston, IL 60208, USA\\
$^{5}$Center for Interdisciplinary Exploration and Research in Astrophysics (CIERA), Northwestern University, 1800 Sherman Ave, Evanston, IL 60201, USA \\
$^{6}$Department of Physics and Earth Science, University of Ferrara, via Saragat 1, I-44122 Ferrara, Italy\\
$^{7}$INFN, Sezione di Ferrara, via Saragat 1, I-44122 Ferrara, Italy \\
$^{8}$INAF, Osservatorio Astronomico d'Abruzzo, via Mentore Maggini snc, 64100 Teramo, Italy \\
$^{9}$MIT-Kavli Institute for Astrophysics and Space Research, 77 Massachusetts Ave., Cambridge, MA 02139, USA \\
$^{10}$Department of Particle Physics and Astrophysics, Weizmann Institute of Science, 234 Herzl St., 7610001 Rehovot, Israel\\
$^{11}$Astrophysics Research Institute, Liverpool John Moores University, 146 Brownlow Hill, Liverpool L3 5RF, UK  \\
$^{12}$Caltech Optical Observatories, California Institute of Technology, Pasadena, CA 91125, USA \\
$^{13}$Division of Physics, Mathematics and Astronomy, California Institute of Technology, Pasadena, CA 91125, USA \\
$^{14}$Institut f\"ur Physik, Humboldt-Universit\"at zu Berlin, D-12489 Berlin, Germany \\
$^{15}$Deutsches Elektronen Synchrotron DESY, Platanenallee 6, D-15738 Zeuthen, Germany \\
$^{16}$The Oskar Klein Centre, Department of Physics, Stockholm University, Albanova University Center, SE 106 91 Stockholm, Sweden \\
$^{17}$The Oskar Klein Centre, Department of Astronomy, Stockholm University, AlbaNova University Center, SE-106 91 Stockholm , Sweden \\
$^{18}$Las Cumbres Observatory, 6740 Cortona Drive, Suite 102, Goleta, CA 93117-5575, USA \\
$^{19}$Department of Astronomy, University of California, Berkeley, CA 94720-3411, USA \\
$^{20}$DIRAC Institute, Department of Astronomy, University of Washington, 3910 15th Avenue NE, Seattle, WA 98195, USA \\
$^{21}$IPAC, California Institute of Technology, 1200 E. California Blvd, Pasadena, CA 91125, USA \\
$^{22}$LPNHE, CNRS/IN2P3, Sorbonne Universit\'e, Universit\'e Paris-Cit\'e, Laboratoire de Physique Nucl\'eaire et de Hautes \'Energies, 75005 Paris, France \\
}

\date{Accepted XXX. Received YYY; in original form ZZZ}

\pubyear{2023}

\begin{document}
\label{firstpage}
\pagerange{\pageref{firstpage}--\pageref{lastpage}}
\maketitle

\begin{abstract}
Early observations of transient explosions can provide vital clues to their progenitor origins. In this paper we present the nearby Type Iax (02cx-like) supernova (SN), SN 2020udy that was discovered within hours ($\sim$7 hr) of estimated first light. An extensive dataset of ultra-violet, optical, and near-infrared observations was obtained, covering out to $\sim$150 d after explosion. SN 2020udy peaked at $-$17.86$\pm$0.43 mag in the \textit{r} band and evolved similarly to other `luminous' SNe Iax, such as SNe 2005hk and 2012Z. Its well-sampled early light curve allows strict limits on companion interaction to be placed. Main-sequence companion stars with masses of 2 and 6 \msun\ are ruled out at all viewing angles, while a helium-star companion is allowed from a narrow range of angles (140--180$^{\circ}$ away from the companion). The spectra and light curves of SN 2020udy are in good agreement with those of the `N5def' deflagration model of a near Chandrasekhar-mass carbon-oxygen white dwarf. However, as has been seen in previous studies of similar luminosity events, SN 2020udy evolves slower than the model. Broad-band linear polarisation measurements taken at and after peak are consistent with no polarisation, in agreement with the predictions of the companion-star configuration from the early light-curve measurements. The host galaxy environment is low metallicity and is consistent with a young stellar population. Overall, we find the most plausible explosion scenario to be the incomplete disruption of a CO white dwarf near the Chandrasekhar-mass limit, with a helium-star companion.

\end{abstract}




\section{Introduction}

In recent years, all-sky transient surveys have discovered new classes of exotic extragalactic transients. In particular, the increased depth and higher cadence have allowed fainter and faster events to be found than previously possible. One such class is Type Iax supernovae (SNe Iax) or 02cx-like after the prototype of this class \citep[][]{2006AJ....132..189J,2007PASP..119..360P}. These are SNe that share some characteristics with normal Type Ia SNe (SNe Ia) but deviate in a number of key ways, such as being fainter at peak at $-12$ to $-$18.5 mag, lacking a secondary maximum in their redder light-curve bands, having lower ejecta velocities around peak at $\sim$3000--7000 \kms\ \citep[e.g.,][]{2017hsn..book..375J}, and alongside forbidden lines in their late-time spectra, permitted \Feii\ lines are also detected \citep{2006AJ....132..189J,2008ApJ...680..580S,2010ApJ...708L..61F,2013ApJ...767...57F,2016MNRAS.461..433F,2014A&A...561A.146S,2015A&A...573A...2S}. They do not follow the Phillips relation between the peak luminosity and decline rate after peak that would allow them to be used as cosmological distance indicators \citep{1993ApJ...413L.105P}. Type Iax SNe are found to occur nearly exclusively in young stellar populations but most likely originate from thermonuclear explosions of white dwarfs \citep[e.g.,][]{2018MNRAS.473.1359L}. 

\cite{2014Natur.512...54M} examined pre-explosion imaging of the nearby SN Iax,  SN 2012Z and identified a luminous blue source in the images, which is consistent with a helium-rich companion star. Very deep very late-time observations of SN 2012Z  at $\sim$1400 d after explosion have still found a source brighter than the pre-explosion source, but the origin is unclear \citep{mccully22}. A companion star containing helium could suggest that signatures of helium would be observable in the near-infrared spectra of SNe Iax, but this has not been detected to limits of $<10^{-3}$ \msun\ in the handful of events with suitable data  \citep{2019A&A...622A.102M}. However, it is suspected that the amount of helium stripped from a helium-rich companion star in a SN Iax is typically low and could be hidden by other features in the late-time spectra \citep{zeng_iax_he}. Radio and x-ray observations of the environment of SN Iax, SN 2014dt put limits on mass-loss rates from the progenitor system that are consistent with those of helium-star companions \citep{2021MNRAS.505.1153S}. The overluminous interacting SN Ia, SN 2020eyj has rare signatures of interaction with helium-rich circumstellar material (CSM), as well as the detection of a radio counterpart. SN 2020eyj is hypothesised to have a helium-rich companion star \citep{kool_he}. 

Several explosion models have been proposed to explain SNe Iax, such as the popular model of the deflagration of a near-Chandrasekhar mass carbon-oxygen (CO) white dwarf, where the deflagration is not sufficiently energetic to unbind the star and so leaves behind a remnant \citep[e.g.][]{2004PASP..116..903B,2012ApJ...761L..23J,2013MNRAS.429.2287K, fink2014}. Their lower luminosities and expansion velocities compared to normal SNe Ia, as well as the characteristics (multiple components of both permitted and forbidden lines) of their late-time spectra can be explained by this scenario \citep{2006AJ....132..189J,2016MNRAS.461..433F,maeda22_iax}. The likely mixing seen in SN Iax ejecta (measured through overlapping velocities of different species) is also consistent with this scenario \citep[e.g.,][]{2020ApJ...902...47M,2022MNRAS.509.3580M}. The flat light curve at late times seen in some SNe Iax is potentially suggestive of a stellar remnant from the failed deflagration \citep{kawabata_2018,2023arXiv230203105C}. 

The polarimetric modelling of the deflagration of a near Chandrasekhar mass white dwarf suggests weak polarisation levels, as well as a weak viewing angle dependence \citep{bulla_iax}. Low levels of overall polarisation were seen in spectra of Type Iax SN 2005hk around maximum light \citep{chornock_05hk,maund_05hk} but the increased polarisation signature at shorter wavelengths (below $\sim$5000 \AA) were suggested to be potentially due to interaction between the ejecta and a main-sequence companion star \citep{bulla_iax}. This deflagration model of a near Chandrasekhar-mass white dwarf does not require a helium companion star.  However, the presence of a helium companion star, as is suggested by the observations discussed above, is consistent with this scenario. A potential progenitor system for this scenario, containing a helium-accreting white dwarf, has recently been identified \citep{greiner_he23}. 

The properties of SNe Iax have also been suggested \citep[e.g.,][]{2015A&A...573A...2S} to be consistent with the pulsational delayed detonation of a white dwarf close to the Chandrasekhar mass \citep{1991A&A...245L..25K,1993A&A...270..223K, 1995ApJ...444..831H, 2014MNRAS.441..532D}. However, these models have relatively layered ejecta, which is at odds with the observed high levels of mixing in SNe Iax. The colour evolution of the models is inconsistent with observations \citep{miller_16fnm}.   The origin of the multiple distinct components of permitted and forbidden lines in the late-time spectra are also more difficult to explain with this model compared to the pure deflagration scenario. 

Deep high-cadence transient surveys have allowed fainter populations of SNe Iax to be uncovered, with the faintest SN Iax \citep[SN2021fcg;][]{2021ApJ...921L...6K} peaking in the \textit{r} band at $-$12.66$\pm$0.20 mag. It has been suggested that the faint end of the SN Iax distribution is most likely from a different progenitor origin than more luminous events, such as deflagrations of Chandrasekhar-mass hybrid carbon-oxygen-neon white dwarfs \citep{kromer2015_hybrid,bravo2016} or the mergers of ONe white dwarfs with black holes or neutron stars \citep{bobrick_22_mergers}. The rate of SNe Iax is estimated at 15$^{+17}_{-9}$ per cent of the SN Ia rate, although the more luminous events (brighter than $-$17.5 mag) are only 0.9$^{+1.1}_{-0.5}$ per cent of the SN Ia rate, implying that the progenitor channel or channels of SNe Iax must be relatively common \citep{srivastav_iaxrates}. Therefore, understanding their origin is essential for understanding the diversity of ways that white dwarfs can explode. 

The canonical `expanding fireball' model, under the assumption of constant temperature and velocity, suggested that the early light curves of SNe Ia should be well-described by a rising power-law with an index of 2 \citep{1999AJ....118.2675R,2006AJ....132.1707C,2010ApJ...712..350H,2011MNRAS.416.2607G,2012ApJ...745...44G}. However, the early light curves of SNe Ia have more recently been shown to depend on the distribution of \nick\ in the ejecta \citep[e.g.~][]{2013ApJ...769...67P,2017MNRAS.472.2787N,magee_ni_org,2020A&A...634A..37M}. Studies of the early light curves of normal SNe Ia have found a mean \textit{r}-band index of 2.01$\pm$0.02 but there are objects that are inconsistent with this value \citep{2020ApJ...902...47M}. The rise of the early light curves of Type Iax SN 2018cxk was found to be close to linear, with an \textit{r}-band index of 0.98$\pm^{0.23}_{0.15}$ \citep{2020ApJ...902...47M}, which is consistent with the `N5def' deflagration model \citep{2013MNRAS.429.2287K, fink2014} predictions of a power-law index of $<1.4$ \citep{2017MNRAS.472.2787N}.

In this paper, we present ultra-violet (UV), optical and near-infrared observations of the nearby SN Iax, SN 2020udy, that was detected by the Zwicky Transient Facility (ZTF) just $\sim$0.3 d ($\sim$7 hr) after the estimated time of first light. In Section \ref{sec:obs}, we present the discovery of SN 2020udy, along with its observations and reduction of the data obtained. In Section \ref{sec:lc_analy}, we present our analysis of the light curves of SN 2020udy, in particular focusing on the very early light curve properties and in Section \ref{sec:spec_analy}, we present the spectroscopic analysis of the optical and near-infrared spectra from peak out to +130 d. In Section \ref{sec:deflag_comp}, a detailed comparison of the observations of SN 2020udy with a specific deflagration model (N5def) of a near Chandrasekhar-mass white dwarf is presented. In Section \ref{sec:discussion}, we discuss SN 2020udy in the context of other SNe Iax, as well as the most promising progenitor models. Finally, we present the conclusions in Section \ref{sec:conclusions}. Throughout this paper we use the AB magnitude system unless otherwise noted.

\section{Observations and data reduction}
\label{sec:obs}

In the following sections, we describe the discovery of SN 2020udy along with details of its host galaxy properties. We also detail the observations and data reduction for the UV and optical photometry, optical and near-infrared spectroscopy, as well as broad-band optical linear polarimetry.

\subsection{Discovery and initial classification}
\label{sec:discovery}
ZTF is an optical transient search operating on the 48-inch (P48) telescope at the Palomar Observatory \citep{bellm19a,Graham2019PASP,Dekany2020PASP}. As part of the ZTF public survey \citep{Bellm2019, fremling_bts}, a new transient ZTF20acdqjeq (hereafter SN 2020udy) was identified by an automated detection algorithm implemented in AMPEL  \citep{NordinAMPEL} on 2020-09-24.35 UT (MJD 59116.35) at 19.63 mag in the \textit{r} band, coincident with the galaxy NGC 0812 at a spectroscopic redshift of $z =$ 0.01722$\pm$0.00003 \citep{1999PASP..111..438F}. The first spectrum of SN 2020udy was obtained at the Palomar 60-in telescope with the Spectral Energy Distribution Machine \cite[SEDM;][]{BlagorodnovaSEDM, RigaultSEDM} on 2020-09-25.41 UT (MJD 59117.41), just 1.06 d after discovery. Initial \textsc{snid} \citep{Blondin2007} spectral template matching produced a best fit with the 02cx-like SN 2005hk at $-$9 d with respect to maximum light at the redshift of the likely host. Given the early detection and low redshift nature of the transient, an extensive follow-up campaign was launched.

\begin{figure}
	\includegraphics[width=9cm]{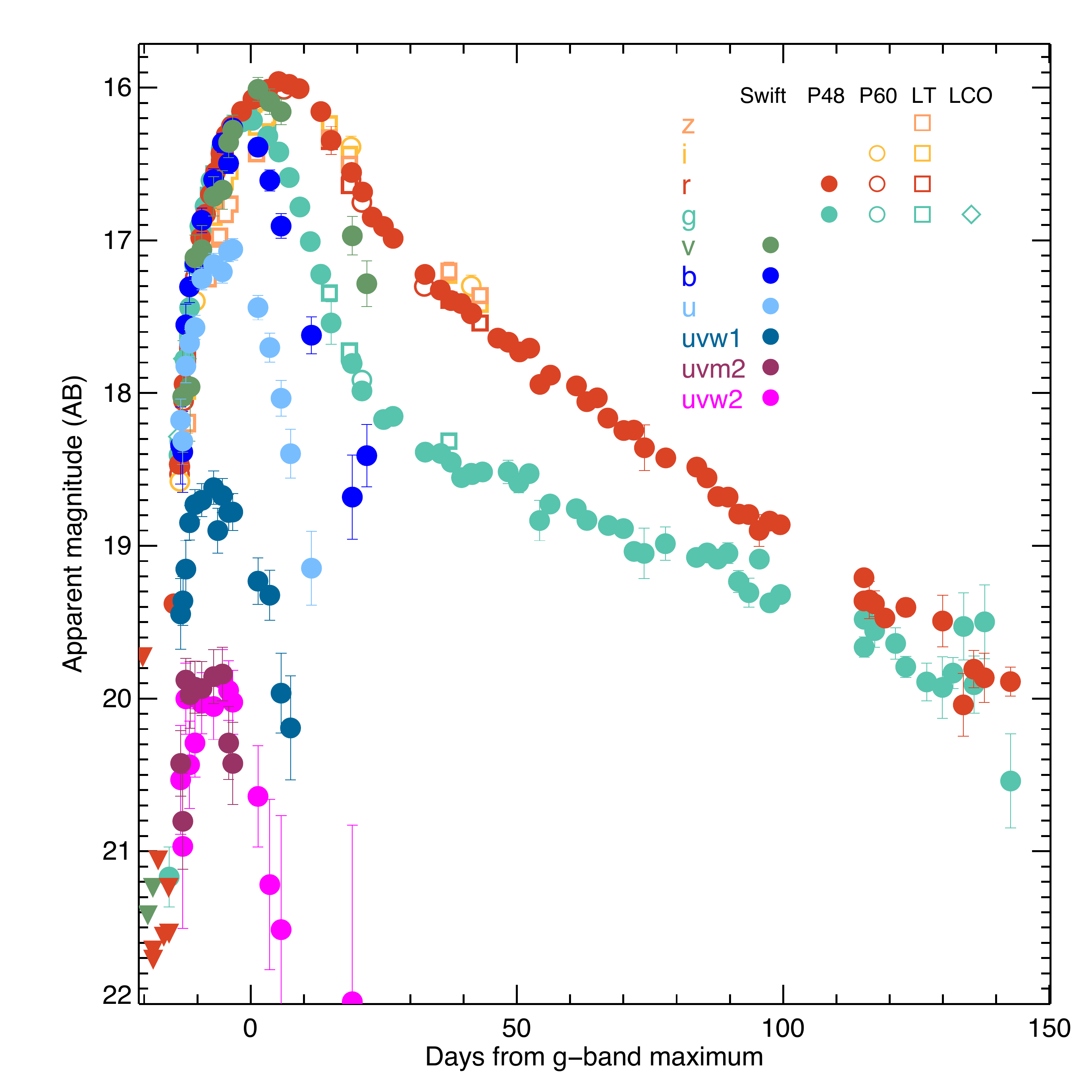}
    \caption{\textit{Swift}, P48, P60, LT, and LCOg light curves of SN 2020udy extending from discovery to $\sim$140 d post \textit{g}-band maximum.  Pre-discovery non-detections in the P48 \textit{gr} bands are shown as downward-facing triangles. The light curves have been corrected for extinction as detailed in Section \ref{sec:discovery}. }
    \label{fig:lc_evo}
\end{figure}

\begin{table}
	\centering
	\caption{Light curve parameters of SN 2020udy and its host galaxy}
	\label{tab:basic_params}
	\begin{tabular}{lc} 
	    \hline
        Parameter & Value \\
		\hline
		Host galaxy & NGC 0812\\
		Redshift & 0.01722$\pm$0.00003\\
		Distance modulus (mag) & 33.83 $\pm$0.43\\
		Milky Way \textit{E(B-V)} (mag) & 0.0658$\pm$0.0004\\
		Epoch of \textit{g}-band first detection (MJD) & 59115.43\\
		Epoch of \textit{r}-band first detection (MJD) & 59116.35\\
		Epoch of \textit{u}-band peak (MJD) & 59127.1$\pm$0.5 \\
		Epoch of \textit{g}-band peak (MJD) & 59131.0$\pm$0.7 \\
		Epoch of \textit{r}-band peak (MJD) & 59136.9$\pm$0.5 \\
		Absolute \textit{u}-band peak (mag) & $-$16.76$\pm$0.44 \\
		Absolute \textit{g}-band peak (mag) & $-$17.58$\pm$0.43 \\
		Absolute \textit{r}-band peak (mag) & $-$17.86$\pm$0.43 \\
		\textit{g-r} colour at \textit{g}-band peak (mag) & 0.12$\pm$0.03 \\
		Epoch of \textit{g}-band first light (MJD) & 59115.1$\pm$0.1 \\
		Epoch of \textit{r}-band first light (MJD) & 59115.2$\pm$0.1\\
		Rest frame \textit{g}-band rise time (d) & 15.6$\pm$0.7 \\
		Rest frame \textit{r}-band rise time (d) & 21.3$\pm$0.5 \\
		Time from first light to \textit{g}-band first detection (hr) & 7.2$\pm$2.4\\
		Early power-law index in the \textit{g}-band & 1.38$\pm$0.11	\\
		Early power-law index in the \textit{r}-band & 1.29$\pm$0.07	\\
		\hline
	\end{tabular}
 \begin{flushleft}
  \end{flushleft}
\end{table}

\subsection{Host galaxy properties}
\label{sec:host}
The host of SN 2020udy is a spiral galaxy, NGC 0812, and SN 2020udy is offset from the centre of the galaxy by 15 kpc at the measured redshift. The distance modulus of the galaxy has been estimated using the Tully-Fisher method to be 33.83$\pm$0.43 mag \citep{2014MNRAS.444..527S}, corresponding to a distance of 58.3 Mpc. We do not apply any host galaxy extinction correction because of the lack of \NaiD\ features in the spectra (see Section \ref{sec:spec_info}). Although the correlation between \NaiD\ strength and related extinction has a significant scatter \citep{Poznanski2012}, the absence of \NaiD\ suggests negligible host extinction, which is consistent with the offset location from the core of the host galaxy. Milky Way extinction of \textit{E(B -- V)} of 0.0658 mag  \citep{SF11} is corrected using the \cite{Fitz99} extinction law with a total-to-selective extinction, $R_{V}$ of 3.1.

The metallicity of the environment of SN 2020udy was measured from a late-time spectrum obtained at +119 d post \textit{g}-band maximum light (see Section \ref{sec:spec_info}). The `N2' method \cite{2004MNRAS.348L..59P} that measures the line flux ratio of \Nii\ to H$\alpha$ was used with the calibration of \cite{2013A&A...559A.114M}. H$\alpha$ is clearly visible in the spectrum but only a limit can be placed for \Nii\ 6583 \AA. This results in a 3 $\sigma$ upper limit to the metallicity of 12 + log(O/H) $<$ 8.30.  Assuming a solar oxygen abundance of 12 + log(O/H)  = 8.69 \citep{2009ARA&A..47..481A}, this is equivalent to $<$0.41 \zsun.

\subsection{Optical and UV photometry}
\label{sec:opt_phot}

The ZTF P48 \textit{gr} images were processed using the ZTF pipeline \citep{Masci2019} and the difference imaging algorithm, ZOGY \citep{Zackay2016}. Photometry was performed at the SN position using the \textsc{fpbot} forced photometry pipeline \citep{ztffps}. An early detection at 5.6 $\sigma$ at 21.41$\pm0.20$ mag in the \textit{g} band was recovered on MJD 59115.43 (0.9 d earlier than the \textit{r}-band point through which the new source was identified).

Photometry was obtained in the \textit{gri} bands using the Palomar 60-inch telescope with the SED Machine \citep{BlagorodnovaSEDM} and reduced using the FPipe pipeline of \cite{Fremling2016}. Photometry was also obtained at the Liverpool Telescope \citep[LT;][]{SteeleLT} with the IO:O imager in the \textit{griz} bands and was reduced using custom pipelines. Two \textit{g}-band points on the rise were also obtained at the Las Cumbres Observatory 1-m array  \citep[LCO;][]{2013PASP..125.1031B} and reduced using the FPipe pipeline. The time of peak in the different bands was estimated using a low-order polynomial fit to the data around peak. The peak of the \textit{g}-band was found to be 59131.0$\pm$0.7  and the \textit{r}-band light curve peaked $\sim$6 d later at MJD = 59136.9$\pm$0.5. 

UV imaging in the \textit{uvw2, uvm2, uvw1, u, B, V} bands was obtained with the Neil Gehrels \textit{Swift} Observatory using the Ultra-violet/Optical Telescope \citep[UVOT;][]{Roming2005} at epochs from $-$14 to +20 d with respect to the \textit{g}-band peak. 
Data were retrieved from the NASA \textit{Swift} Data Archive\footnote{\url{https://heasarc.gsfc.nasa.gov/cgi-bin/W3Browse/swift.pl}} and processed using the UVOT data analysis software HEASoft version 6.30.1\footnote{ \url{https://heasarc.gsfc.nasa.gov/}}. Source counts were extracted from the images using a region of $5''$. The background was estimated using a significantly larger region outside of the host galaxy. The count rates were obtained from the images using the \textit{Swift} tool uvotsource. They were converted to magnitudes using the UVOT photometric zero points \citep{Breeveld2011} and the latest calibration files from February 2022. UVOT reference images were obtained between 2021-03-07 and 2021-03-11, more than six months after the SN discovery and subtracted from the early images to remove the host contribution. The UV bands were seen to peak even earlier than the optical at $-$3.9$\pm$0.9, $-$5.4$\pm$1.0 and $-$6.0$\pm$1.0 d, in the rest-frame with respect to \textit{g}-band peak, in the \textit{u}-, \textit{uvw1}- and \textit{uvw2}-bands, respectively.

At the estimated distance of the SN, the corresponding peak \textit{g}- and \textit{r}-band absolute magnitudes are $-$17.58$\pm$0.43 and $-$17.86$\pm$0.43 mag, respectively.    The UV and optical light curves of SN 2020udy are shown in Fig.~\ref{fig:lc_evo} relative to \textit{g}-band peak. An analysis of the early phases of the light curves, as well as a comparisons to other SNe Iax, is presented in Section \ref{sec:lc_analy}. The post-peak evolution of SN 2020udy and other SNe Iax is different to those of normal SNe Ia because SNe Iax do not show a secondary peak in the redder bands at a few weeks past peak.

\begin{figure}
	\includegraphics[width=\columnwidth]{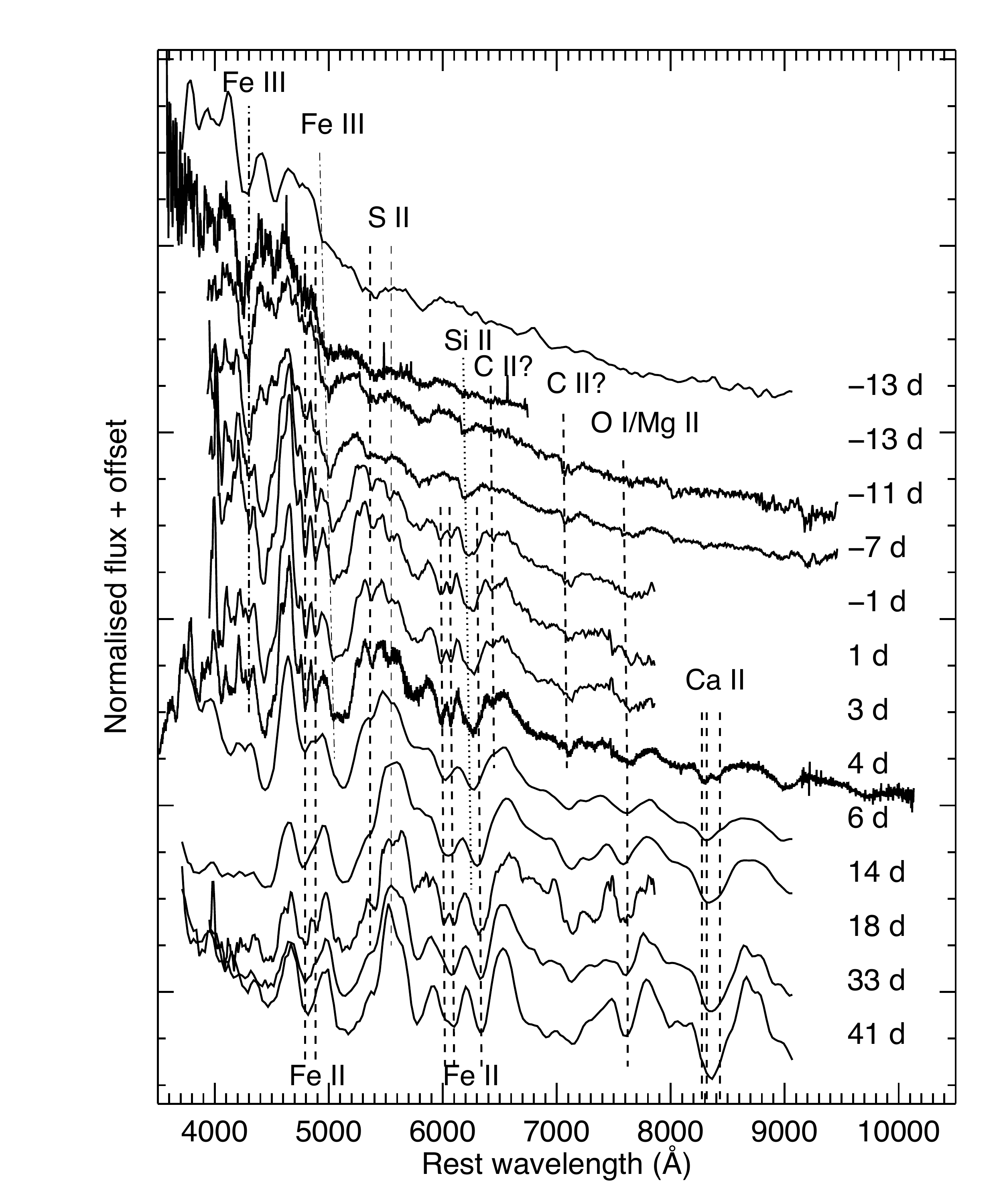}
    \caption{The optical spectral evolution of SN 2020udy at phases from $-$13 to +41 d relative to \textit{g}-band maximum. The velocity evolution of two features of Fe III (dash-dotted) and Si II (dotted) are marked. Prominent absorption features of other features are marked as vertical dashed lines. }
    \label{fig:opt_spec}
\end{figure}

\subsection{Optical and near-infrared spectroscopy}
\label{sec:spec_info}

\begin{table}
	\centering
	\caption{Information on the optical and near-infrared spectroscopic observations of SN 2020udy.}
	\label{tab:spec_obs}
	\begin{tabular}{lccc} 
		\hline
		Night of observation & MJD$^a$ & Phase (d)$^b$ & Telescope+instrument \\
		\hline
		20200925 & 59117.4 & $-$13.4 &P60+SEDM \\
		20200925 & 59117.5 & $-$13.3 &GeminiN+GMOS \\
		20200927 & 59119.9 & $-$10.9 & NOT+ALFOSC \\
		20201002 & 59124.0 & $-$6.9 & NOT+ALFOSC \\
		20201003 &59125.3&$-$5.6&P200+TSPEC\\
		20201008 & 59130.1 & $-$0.9 & LT+SPRAT \\
		20201010 & 59132.2 & +1.2 & LT+SPRAT \\
		20201012 & 59134.1 & +3.0 & LT+SPRAT \\
		20201012 & 59134.6 & +3.5 & Keck+LRIS \\
		20201015 & 59137.4 & +6.3 & P60+SEDM \\
		20201015 & 59137.6  & +6.5 & IRTF+SpeX \\
		20201023 & 59145.3 & +14.1 & P60+SEDM \\
		20201027 & 59149.1 & +17.8 & LT+SPRAT \\
        20201111 & 59173.1 & +32.6 & P60+SEDM \\
        20201116 & 59252.1 & +41.4 & P60+SEDM \\
		20210207 & 59252.1 & +119.0 & MMT+Binospec \\ 
		20210221 & 59267.2 &+133.9&Keck+NIRES \\
        20200224 &59269.8 &+136.5 & NOT+ALFOSC \\   
		\hline
	\end{tabular}
 \begin{flushleft}
$^a$MJD = Modified Julian date.\\
$^b$Rest frame phase relative to \textit{g}-band maximum light of 59131.0$\pm$0.7.\\
 \end{flushleft}
\end{table} 

\begin{table*}
	\centering
	\caption{Information on the broad-band linear polarimetry of SN 2020udy taken at the NOT+ALFOSC.}
	\label{tab:pola_obs}
	\begin{tabular}{lcccccccccc} %
		\hline
		Night of observation & MJD$^a$ & Phase (d)$^b$ & Band& Stokes \textit{Q} (\%)$^c$  & Stokes \textit{U} (\%)$^c$  & Polarisation degree (\%)$^d$ & Intrinsic polarisation degree (\%)$^e$   \\
		\hline
20201009 &59131.1 & 0.1  &\textit{B}  &  $-$0.42$\pm$0.15  &   0.21$\pm$0.15   &  0.47$\pm$0.15  &  0.23$\pm$0.16  \\
20201009 &59131.1&  0.1 &\textit{V} &$-$0.66$\pm$0.12  &   0.52$\pm$0.12  &  0.84$\pm$0.12  & 0.04$\pm$0.12  \\
20201016 &59138.1 & 7.0  &\textit{B}  &  $-$0.91$\pm$0.19  &   0.13$\pm$0.16   &  0.92$\pm$0.19  &  0.31$\pm$0.18  \\
20201016 &59138.1  & 7.0& \textit{V} &$-$0.63$\pm$0.13  &   0.56$\pm$0.15  &  0.84$\pm$0.14  & 0.05$\pm$0.14  \\
20201109 &59162.1&30.6& \textit{V} &$-$0.91$\pm$0.19  &   0.51$\pm$0.19  &  1.04$\pm$0.19  & 0.26$\pm$0.19  \\
 \hline
	\end{tabular}
 \begin{flushleft}
$^a$MJD = Modified Julian date.\\
$^b$Rest frame phase relative to $g$-band maximum light of 59131.0$\pm$0.7.\\
$^c$These are the measured Stokes parameters, including contribution from the ISP. \\
$^d$Computed as $P = \sqrt{Q^2 + U^2}$. \\
$^e$Intrinsic polarisation degree, computed as $P = \sqrt{(Q - Q_{ISP})^2 + (U - U_{ISP})^2}$ and further corrected for polarisation bias following \cite{Plaszczynski14}. The estimated ISP was $Q_{ISP} = (-0.59 \pm 0.02)$\% and $U_{ISP} = (0.49 \pm 0.02)$\% in the $V$-band and $Q_{ISP} = (-0.64 \pm 0.05)$\% and $U_{ISP} = (0.37 \pm 0.04)$\% in the $B$-band.\\ 
\end{flushleft}
\end{table*}

Optical and near-infrared photospheric-phase spectra were obtained at phases ranging from $-$13 to +41 d relative to \textit{g}-band maximum using the telescopes and instruments listed
in Tables \ref{tab:spec_obs} and \ref{tab:spec_obs_add}\footnote{The additional spectra in Table \ref{tab:spec_obs_add} are not discussed further because they are lower signal-to-noise (S/N) than spectra obtained at very similar phases. However, they are available from WISeREP.}. Three nebular phase spectra were obtained, two in the optical and one in the near-infrared. The P60+SED Machine spectra were reduced using \textsc{pysedm}\footnote{\url{https://github.com/MickaelRigault/pysedm}} \citep{RigaultSEDM,Kim2022PASP}. We observed SN 2020udy with the Gemini Multi-Object Spectrographs (GMOS, \citealp{2004PASP..116..425H}) at the Gemini-North observatory (Programme: GN-2020B-Q-901, PI: Gal-Yam).  The GMOS data were reduced following standard routines using the Gemini \textsc{iraf}\footnote{{IRAF} is distributed by the National Optical Astronomy Observatories, which are operated by the Association of Universities for Research in Astronomy, Inc., under cooperative agreement with the National Science Foundation.} package.  Three spectra were obtained with Alhambra Faint Object Spectrograph and Camera (ALFOSC\footnote{\url{http://www.not.iac.es/instruments/alfosc}}) at the Nordic Optical Telescope (NOT) and were reduced using standard reduction techniques in \texttt{pypeit} \citep{pypeit:joss_arXiv}. Spectra were obtained with LT and the SPectrograph for the Rapid Acquisition of Transients \citep[SPRAT;][]{Piascik2014SPIE} and were reduced using the automated LT pipeline \citep{2012AN....333..101B}. One spectrum was obtained with the Low-Resolution Imaging Spectrometer \cite[LRIS;][]{1995PASP..107..375O} on the Keck I 10-m telescope and reduced using \texttt{lpipe} \citep{perley_lpipe}. One spectrum was obtained with Binospec \citep{binospec} on the 6.5-m MMT telescope and reduced using the standard Binospec pipeline. The photospheric optical spectral sequence of the spectra detailed in Table \ref{tab:spec_obs} are shown in Fig.~\ref{fig:opt_spec}, with the main spectral features identified.  

We obtained a near-infrared spectrum of SN 2020udy with the Palomar 200-inch telescope using TripleSpec \citep{herter08} at $-$6 d with respect to maximum light and a second spectrum using SpeX on the NASA Infrared Telescope Facility (\citealt{Rayner2003}; Programme: 2020B087, PI: De) at +6 d after maximum light. Both spectra were reduced using the \texttt{spextool} pipeline \citep{Cushing2004}. The extracted spectra were flux calibrated and corrected for telluric absorption with standard star observations using the \texttt{xtellcor} package  \citep{Vacca2003}. A final near-infrared spectrum was obtained at 133.9 d with respect to maximum light using NIRES on the Keck Telescope and was reduced using \texttt{pypeit} \citep{pypeit:joss_arXiv} and standard reduction procedures.

\subsection{Broad-band linear polarimetry}

Broad-band linear polarimetry was obtained with ALFOSC at the NOT on three epochs between \textit{g}-band maximum and $+$30 d. Observations were obtained in the \textit{B}  and \textit{V} bands, except the last epoch, where only \textit{V} was used (see Table \ref{tab:pola_obs}). 
All observations were obtained at 4 half-wave plate retarder angles (0$^\circ$, 22.5$^\circ$, 45$^\circ$ and 67.5$^\circ$) and the data were reduced and analysed following the methods outlined in \cite{2023arXiv230108111P}. These authors presented a study of a sample of superluminous SNe, and investigated in detail NOT/ALFOSC polarimetry data including issues such as the shape of the point spread function (PSF), the optimal choice of aperture, the impact of S/N, the effect of the moon and the use of field stars to determine the (Galactic) interstellar polarisation (ISP).  The data of SN~2020udy were reduced with the same custom pipeline as in \cite{2023arXiv230108111P}. We measured the Stokes parameters (\textit{Q}, \textit{U}) for both SN~2020udy and a comparison star of adequate S/N in the same field of view. The comparison star was used to estimate the Galactic ISP along the line of sight, and the ISP was removed vectorially  from the Stokes parameters of SN~2020udy to obtain the transient intrinsic polarisation. Including more stars of lower S/N proved consistent but not constraining for the Galactic ISP. 
The SN measurements were corrected for polarisation bias following \cite{Plaszczynski14}. Our measurements are included in Table \ref{tab:pola_obs}.

\section{Light curve analysis}
\label{sec:lc_analy}

 In Section \ref{sec:lc_para}, we discuss the optical and UV light curves and colour curves of SN 2020udy in the context of other SNe Iax, as well as its position in the peak absolute against rise time parameter space of SNe Iax and normal SNe Ia. We investigate the novel early light-curve data of SN 2020udy and perform a power-law fit to these data (Section \ref{sec:early_lc}).
 
\begin{figure*}
	\includegraphics[width=17cm]{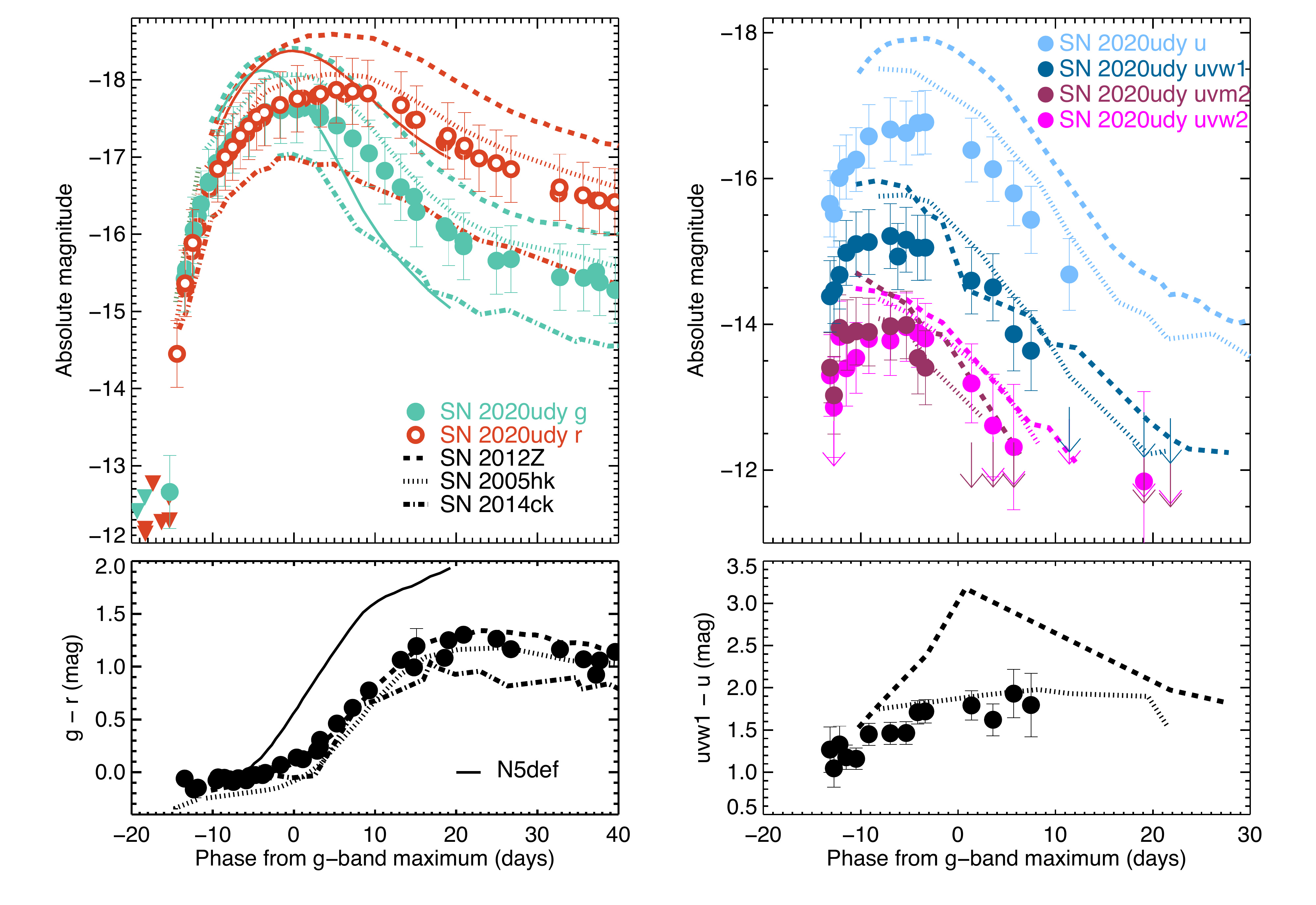}
    \caption{The top left panel shows the rest-frame light curves in the \textit{g} and \textit{r} bands for SN 2020udy compared to three 02cx-like events with good pre-maximum coverage, SNe 2005hk (dotted line), 2012Z (dashed line), and 2014ck (dash-dotted line). The \textit{g} and \textit{r}-band N5def model light curves of \protect \cite{2013MNRAS.429.2287K} are also shown as solid lines. The top right panel shows the rest-frame light curves in the \textit{Swift} \textit{uvw1,uvm2,uvw2} bands for SN 2020udy compared to those of two 02cx-like events with early \textit{Swift} coverage, SNe 2005hk (dotted line) and 2012Z (dashed line). The bottom left panel shows the \textit{g--r} band colour curves compared to SNe 2005hk, 2012Z and 2014ck along with the N5def colour curve. The bottom right panel shows the \textit{uvw1 - u} band colour curve for SN 2020udy along with those of SN 2005hk and SN 2012Z.  }
    \label{fig:lc_comp}
\end{figure*}

\begin{figure}
	\includegraphics[width=\columnwidth]{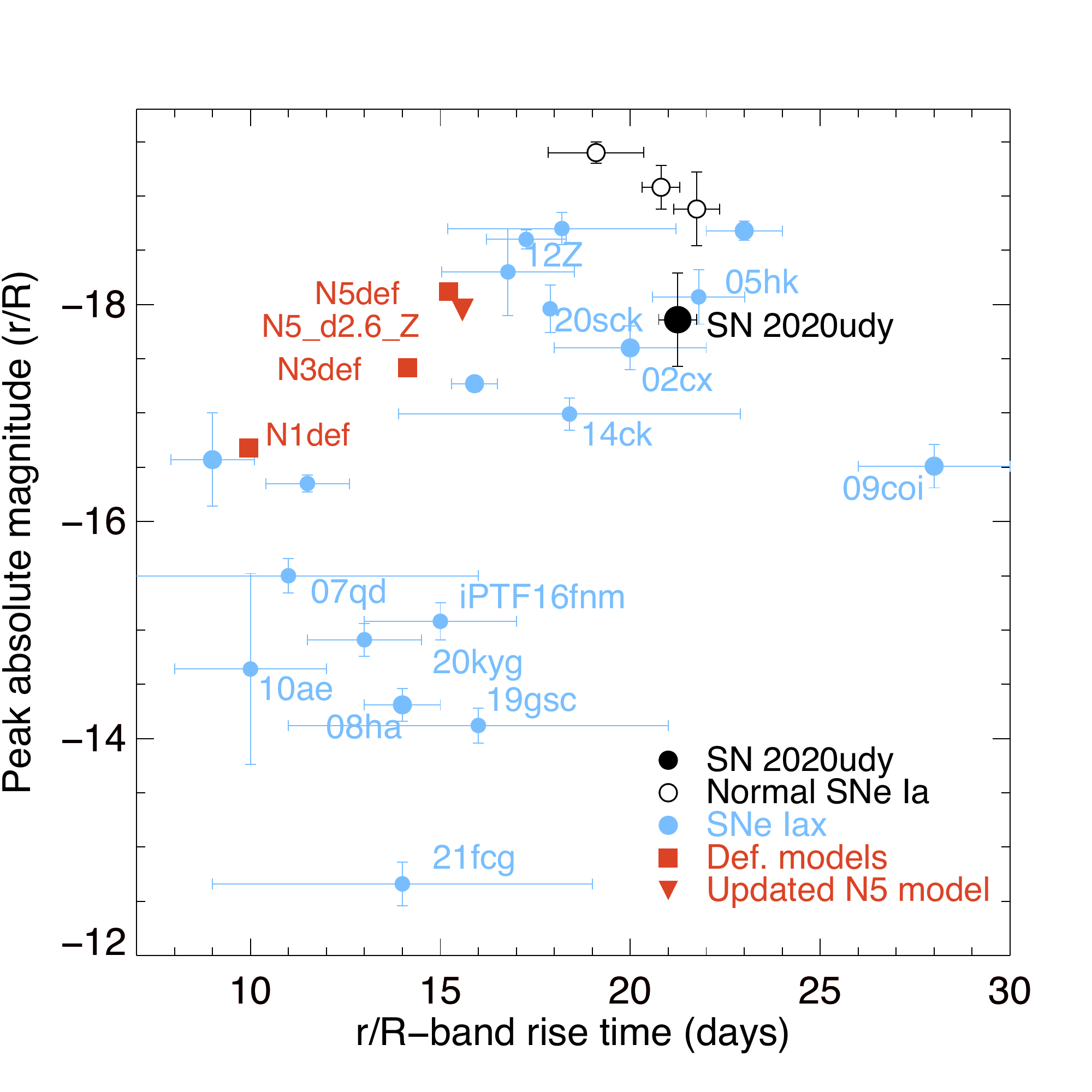}
    \caption{Absolute magnitude against rise time in the \textit{r/R} band of SN 2020udy (solid black) compared to a literature sample of SNe Iax (blue) and normal SNe Ia (open black). The original deflagration models of \protect \cite{2013MNRAS.429.2287K} are shown as red squares and the updated N5 model (N5\_d2.6\_Z) of \protect \cite{lach22} is shown as a red triangle.  }
    \label{fig:absmag_rise}
\end{figure}

\begin{figure*}
	\includegraphics[width=12cm]{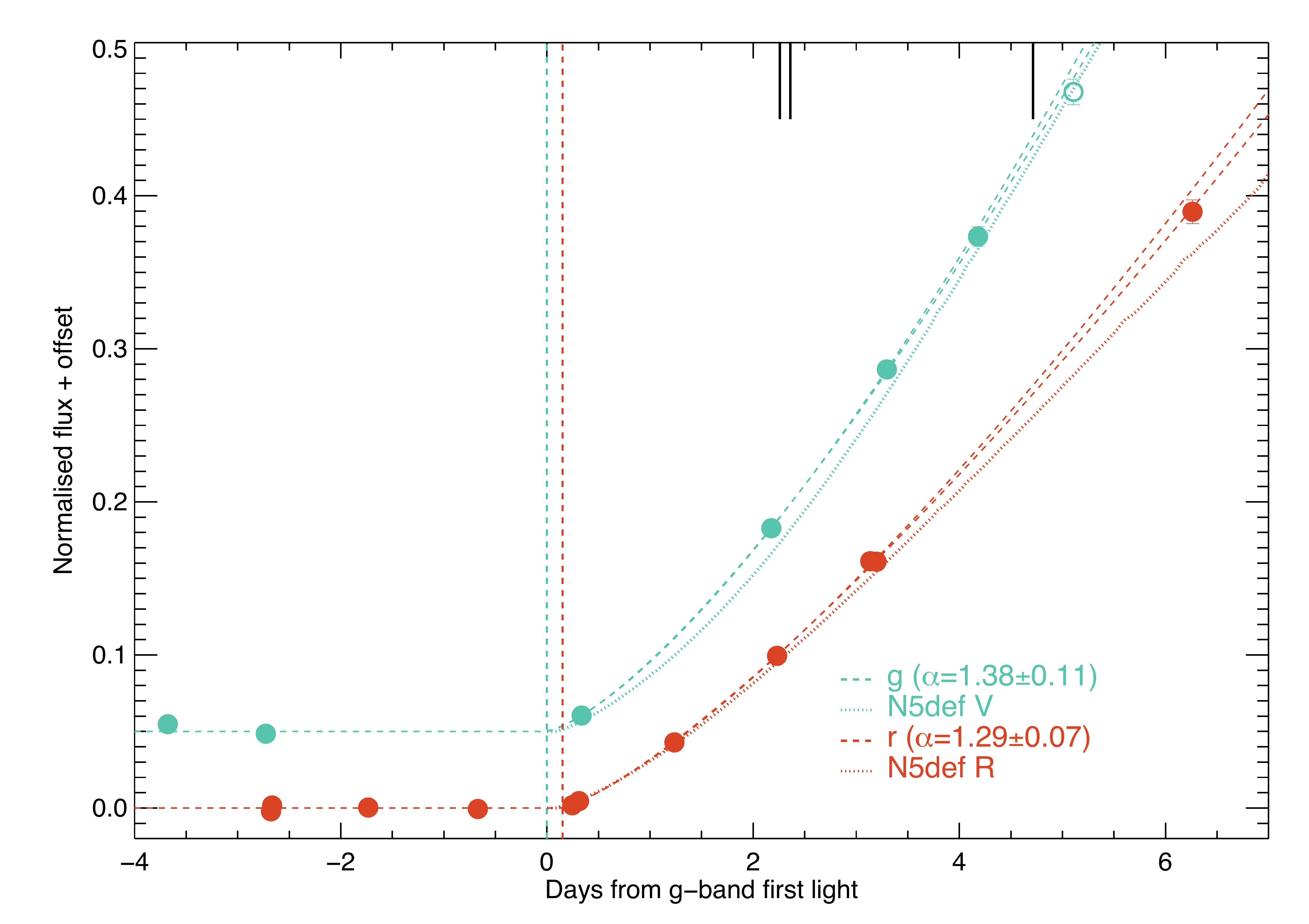}
    \caption{The early rest-frame \textit{g}- and \textit{r}-band light curves of SN 2020udy entering the power law fits are shown as solid data points. The light curves have been normalised to peak flux. The two best-fitting power laws (varying the data points going into the fits) are shown as dashed lines for each band, with the parameters of the power-law index given in the legend. The open circle for the \textit{g}-band is not included in the fitting as it exceeds the flux cut-off of 40 per cent with respect to the peak flux. The \textit{g} band data and its fits are offset by 0.05 for clarity. The N5def \textit{V} and \textit{R} band light curves of \protect \cite{2017MNRAS.472.2787N} are also shown as dotted lines. The vertical solid black lines along the top of the plot show the spectral epochs and the vertical dashed lines show the the estimated time of \textit{g}-band (green) and \textit{r}-band first light. }
    \label{fig:early_flux}
\end{figure*}

\subsection{Comparing SN 2020udy light curve properties to other SNe Iax}
\label{sec:lc_para}
The top left panel of Fig.~\ref{fig:lc_comp} shows a comparison between the absolute \textit{gr}-band magnitude light curves of SN 2020udy and three other SNe Iax that have similar peak luminosities, SNe 2005hk \citep{2008AJ....136.2306H}, 2012Z \citep{2015A&A...573A...2S}, and 2014ck \citep{2016MNRAS.459.1018T}. The light curve of SN 2020udy begins much earlier (within 0.3$\pm$0.1 d of estimated first light, see Section \ref{sec:early_lc}) than the comparison objects but on the rise, at peak, and post peak, SN 2020udy follows a similar evolution to these events. The \textit{g} band peaks before the \textit{r} band in all the objects, as was seen for SN 2020udy (Section \ref{sec:opt_phot}), with a lag of $\sim$6 d of the \textit{r}- relative to \textit{g}-band. The light curve evolution of SN 2020udy is most similar to that of SN 2005hk (dashed lines in Fig.~\ref{fig:lc_comp}). 

The \textit{gr}-band light curves of the N5def deflagration model of a Chandrasekhar-mass white dwarf \citep{2013MNRAS.429.2287K,fink2014} are also shown for comparison. The \textit{g} and \textit{r}-band magnitudes were estimated from the model spectra, instead of the early light curve models of \cite{2017MNRAS.472.2787N}, to ensure coverage over a later phase range for comparison with the other events -- the light curve models only extend to +10 d from explosion.  The early light-curve comparison of SN 2020udy to the N5def models is presented in Section \ref{sec:early_lc}. 

The top right panel of Fig.~\ref{fig:lc_comp} shows the \textit{Swift} \textit{u}, \textit{uvw1} and \textit{uvw2} light curves of SN 2020udy compared to those of SN 2005hk \citep{2014Ap&SS.354...89B} and  SN 2012Z \citep{2015A&A...573A...2S}. SN 2020udy appears to be fainter in all the UV bands than SN 2005hk and SN 2012Z. As discussed in Section \ref{sec:host}, we estimated a negligible host galaxy extinction from the absence of \NaiD\ absorption in the spectra and its offset location of 15 kpc from the host galaxy centre so unaccounted for extinction is unlikely to be the cause of these differences. The N5def models are faster evolving light curves, with both the \textit{g} and \textit{r} model light curves peaking earlier than SN 2020udy and the comparison events, as has been identified in previous comparisons between the models and SNe Iax.

The bottom panels show the \textit{g -- r} and \textit{uvw1 -- u} colour curves of SN 2020udy compared to those of the comparison SNe Iax. In the optical bands, SN 2020udy is seen to be most similar to SN 2005hk, with SN 2012Z being brighter in both the \textit{g}- and \textit{r}-bands but with a similar evolution. SN 2014ck is fainter with a steeper decline in the \textit{g}-band compared to SN 2020udy.  The other SNe have little pre-peak data for comparison but after peak the decline rate of SN 2020udy is similar. The \textit{g} -- {r} colour evolution of the N5def deflagration model (estimated from the model spectra of \cite{2013MNRAS.429.2287K}) is shown and it evolves more quickly to the red than SN 2020udy and the comparison SNe Iax. 

Figure \ref{fig:absmag_rise} shows a comparison between the peak absolute \textit{r/R}-band magnitude and rise time of SN 2020udy and some SNe Ia and SNe Iax. The rise time for SN 2020udy is measured using power-law fits to the early light curves (Section \ref{sec:early_lc}).  The majority of the literature sample is from \cite{2016A&A...589A..89M}, with additional points for SN 2014ck \citep{2016MNRAS.459.1018T}, iPTF16fnm \citep{miller_16fnm}, SN 2019gsc \citep{2020ApJ...892L..24S,2020MNRAS.496.1132T}, SN 2020kyg  \citep{srivastav_iaxrates}, SN 2020sck \citep{2022ApJ...925..217D}, and SN 2021fcg \citep{2021ApJ...921L...6K}. Three deflagration models (N1def, N3def, N5def) that provide the closest matches to the brighter end of the SNe Iax are also shown \citep{2013MNRAS.429.2287K,fink2014}, as well as an updated version of N5def called `N5$\_$d2.6$\_$Z' that has very similar output values to the original N5def \citep{lach22}. 

The peak magnitude of SN 2020udy is most similar to SN 2005hk and the rise times of the bulk of SNe Iax with similar luminosities are up to $\sim$50 per cent longer than the N5def deflagration model light curves. SN 2005hk has an estimated \nick\ mass of $\sim$0.2 \msun\ \citep{2007PASP..119..360P}. Given the similarity in luminosity and colour evolution in the optical and UV bands\footnote{SN 2020udy is slightly fainter than SN 2005hk in the optical and near-UV but has a similar brightness in the bluest \textit{Swift} UV filters.} of SN 2005hk to SN 2020udy, we assume that SN 2020udy has a similar \nick\ mass, which is confirmed by the reasonable match to the N5def deflagration light curve model \cite{2013MNRAS.429.2287K} in Section \ref{def_lc}.

\subsection{Constraining the early light curve}
\label{sec:early_lc}

The early light curves of SNe Iax have not be well constrained to date due to the lack of early data. For the Type Iax, SN 2018cxk, an \textit{r}-band power-law index (${\alpha}_r$) of 0.98$\pm^{0.23}_{0.15}$ was measured from its early light curves \citep{2020ApJ...902...47M}, suggestive of a well-mixed ejecta \citep{2017MNRAS.472.2787N}. Samples of normal SNe Ia have ${\alpha}_r$ values with a mean of $\sim$2 \citep{2020ApJ...902...47M} but a diversity of values is seen, in agreement with a range of \nick\ distributions predicted for SNe Ia \citep[e.g.~][]{2013ApJ...769...67P,2017MNRAS.472.2787N,magee_ni_org,2020A&A...634A..37M}.  

The early ZTF \textit{g}- and \textit{r}-band light curves of SN 2020udy were fitted in flux space, with a power law with a free power-law index ${\alpha}$  as detailed in \cite{2020ApJ...902...47M}.  The fit included a Heaviside function, where the flux is equal to zero for time, \textit{t}, before the time of first light, \textit{t}$_{\mathrm{fl}}$, e.g.,~t\textit{t} $<$ t$_{\mathrm{fl}}$ and is equal to one for \textit{t} $\geq$ \textit{t}$_{\mathrm{fl}}$. The \textit{g}- and \textit{r}-band light curves were fit independently since the first flux does not have to emerge simultaneously in both bands. The data were initially fit up to 0.4 times the peak flux in each band as has been done in previous studies \cite[e.g.][]{2020ApJ...902...47M}. To account for potential systematics due to the epochs of data used in the fit, the fitting was performed including/excluding the nearest light-curve epoch to the 40 per cent flux cutoff. The results of the different fits were found to be consistent, i.e.,~$\leq$0.04 difference in the power-law indices for the fits using different data points, which is within the statistical uncertainties on the fits. We took the best-fitting results as the mean of the fits with and without the point near the 40 per cent cutoff (shown as dashed lines in Fig.~\ref{fig:early_flux}), along with the larger uncertainty value\footnote{A simple propagation of uncertainties could not be used due to the high degree of correlation between the data used in the fits so we conservatively chose the larger uncertainty.}. The light curves in Fig.\ref{fig:early_flux} are shown normalised to peak brightness.

The best fitting power-law index in the \textit{g} band, $\alpha_{\mathrm{g}}$, is 1.38$\pm$0.11 and in the \textit{r}-band, $\alpha_{\mathrm{r}}$ is 1.29$\pm$0.07.  These $\alpha$ values are smaller than the mean $\alpha_{\mathrm{r}}$ of  2.01$\pm$0.02 measured for normal SNe Ia \citep{2020ApJ...902...47M}. We found that the best-fitting t$_{\mathrm{fl}}$ for the \textit{g} band is $-$15.6$\pm$0.7 d in the rest frame with respect to peak magnitude in the \textit{g}-band, which is $\sim$0.1 d earlier than that of the \textit{r}-band t$_{\mathrm{fl}}$ relative to \textit{g}-band peak. The rise time from \textit{r}-band t$_{\mathrm{fl}}$ to \textit{r}-band peak is 21.3$\pm$0.5 d.  The first data point in the \textit{g}-band at 5.6$\sigma$ is just  0.3$\pm$0.1 d (7.2$\pm$2.4 hr) in the rest frame after the estimated time of t$_{\mathrm{fl}}$ in the \textit{g}-band, very narrowly beating the SN Iax, SN 2018cxk (ZTF18abclfee) that had an estimated t$_{\mathrm{fl}}$ $\sim$8 hr before the first detection \citep{2020ApJ...902...47M}.  

In Fig.~\ref{fig:early_flux}, we also show the light curves of the Chandrasekhar mass deflagration model, N5def \citep{2013MNRAS.429.2287K, fink2014}, calculated by \cite{2017MNRAS.472.2787N} up to $+$10 d post explosion for the \textit{VR} bands and normalised to peak brightness. The epoch of explosion of the model is assumed to be the same as the \textit{g}-band time of first light, since these models have typical `dark phases' between explosion and first light of $<$0.2 d  \citep{2017MNRAS.472.2787N}, so this is likely a reasonable assumption. The `dark phase' refers to the time between the explosion of the star occurring and the first emergence of light that depends on the distribution of \nick\ in the ejecta.  

These early light curve simulations were performed with the radiative hydrodynamics code \textsc{stella} because it performed better, compared to Monte-Carlo based codes such as \textsc{artis}, in the optically thick conditions present at these phases \citep[see discussion in][]{2017MNRAS.472.2787N}.  There is good agreement between the observed and N5def model early-time light curves but we leave further discussion of this comparison to Section \ref{sec:deflag_comp}. 

One caveat with the comparison between the model and observed light curves is that the model light curves were produced in the Bessell \textit{UBVR} bands in the AB-magnitude system that are different to the observed ZTF\textit{g} and ZTF\textit{r} bands. To estimate the impact of these filter differences on the comparison, we have used the earliest spectrum of SN 2020udy to perform spectrophotometric comparisons of the Bessell \textit{V} and ZTF\textit{g} bands and the Bessell \textit{R} and ZTF\textit{r} bands\footnote{We have used the ZTF\textit{g} and ZTF\textit{r} filter responses from \url{http://svo2.cab.inta-csic.es/svo/theory/fps/index.php?gname=Palomar&gname2=ZTF}}.   The differences are negligible for the \textit{R/r} bands, with the Bessell \textit{R} filter just 0.02 mag fainter than ZTF\textit{r}. For the \textit{V/g} bands, the difference is larger with the Bessell \textit{V} being 0.07 mag fainter than the ZTF\textit{g} band.

\subsection{Broad-band linear polarisation analysis}

\begin{figure}
	\includegraphics[width=\columnwidth]{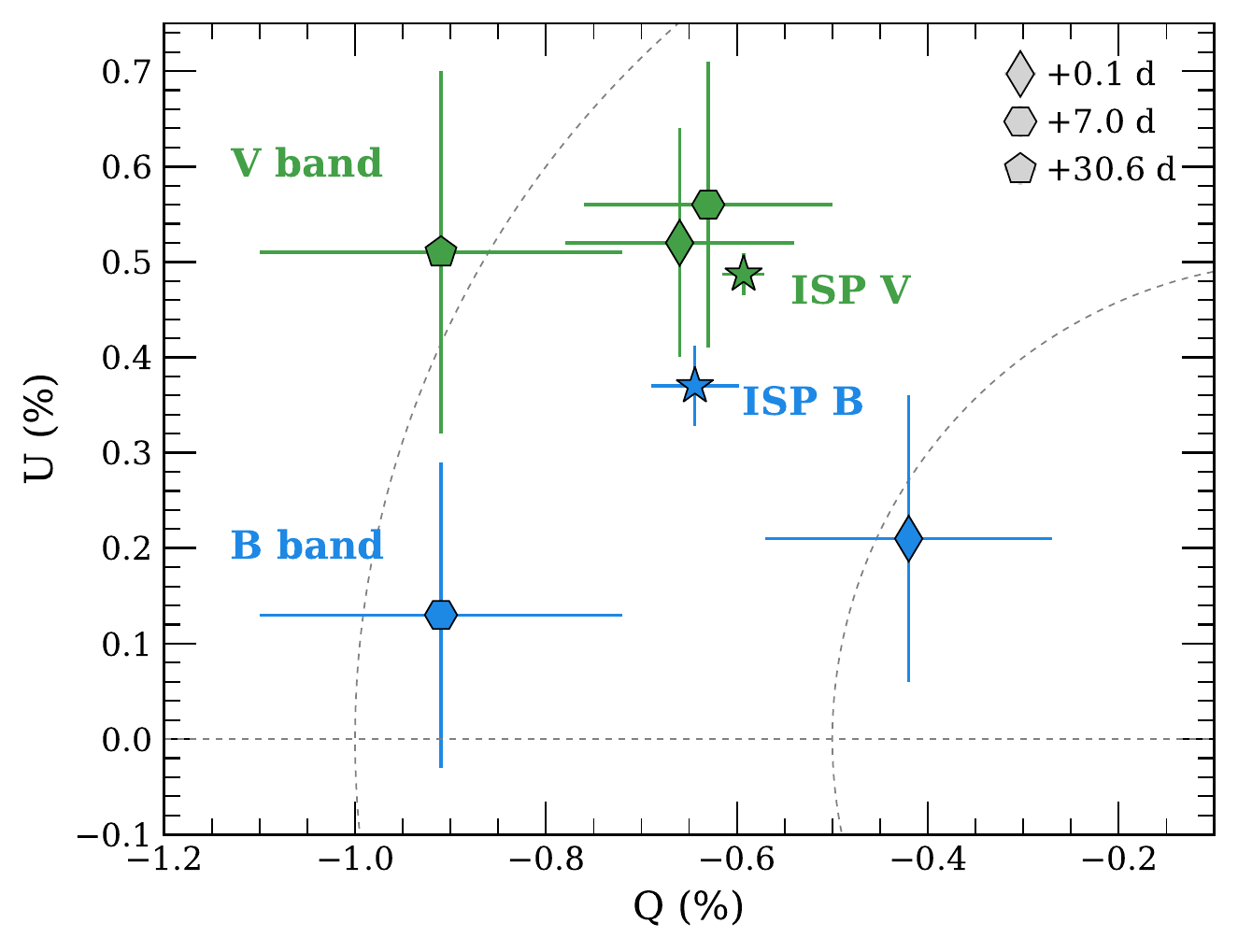}
    \caption{The Stokes \textit{Q - U} plane for three epochs of SN 2020udy in the \textit{B} (blue points) and \textit{V} (green points) bands. The markers indicate the rest-frame phases, as indicated in the legend. The estimated ISP values for both bands are also shown as stars. The grey dashed lines mark \textit{U} = 0 and \textit{P} = 0.5, 1.0 per cent. We emphasize that the data points for SN 2020udy are the observed Stokes parameters. After removal of the ISP, the intrinsic polarization of SN~2020udy is very low (see Table \ref{tab:pola_obs}). }
    \label{fig:polar}
\end{figure}

The intrinsic polarization degree of SN~2020udy is very low and ranges between 0.0--0.3 per cent with typical uncertainties of 0.15 per cent, where these values have been corrected for ISP and polarization bias (Table \ref{tab:pola_obs}).
Furthermore, the location of SN~2020udy on the Stokes plane is in all epochs and filters consistent with the foreground star (Fig.~\ref{fig:polar}). Therefore, we conclude that the SN is not significantly polarised. 

\section{Spectroscopic analysis}
\label{sec:spec_analy}

In this section, we first compare the early phase (pre-peak) spectra of SN 2020udy to a sample of similar luminosity SNe Iax. Then we present the velocity evolution of the prominent lines of \SiII\ and \Feii\ seen in the optical, as well as the detection of likely \Cii\ features (Section \ref{sec:early_phase}). The two early-time near-infrared spectra of SN 2020udy, and the search for \Ci\ and \Hei\ features in the near-infrared, are described in Section \ref{sec:nir_lineids}. In Section \ref{latephase_spec}, we present the optical and near-infrared nebular phase spectra of SN 2020udy. 

\begin{figure*}
	\includegraphics[width=18cm]{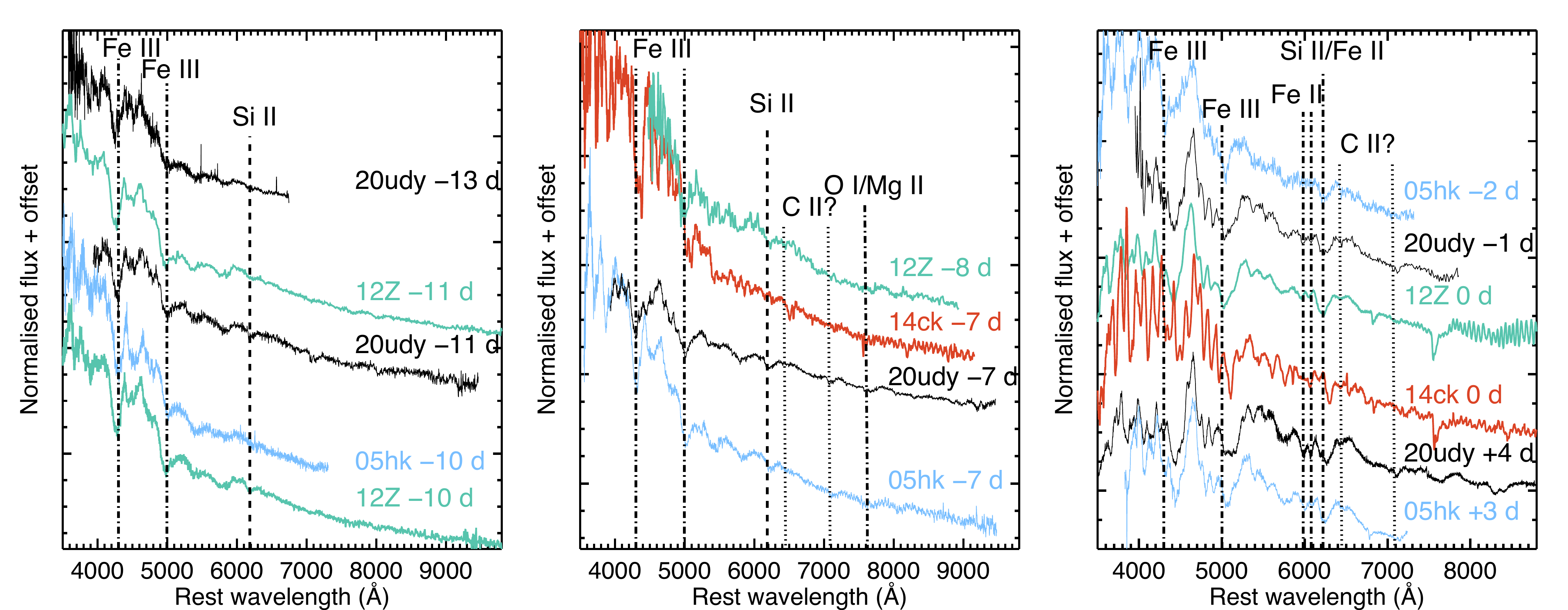}
    \caption{Optical spectra of SN 2020udy (black) compared to those of three other 02cx-like events at similar phases (SNe 2005hk, 2012Z, 2014ck). The date of the spectra with respect to rest frame \textit{g}-band maximum light are given in each panel.}
    \label{fig:opt_spec_comp}
\end{figure*}

\begin{figure}
	\includegraphics[width=\columnwidth]{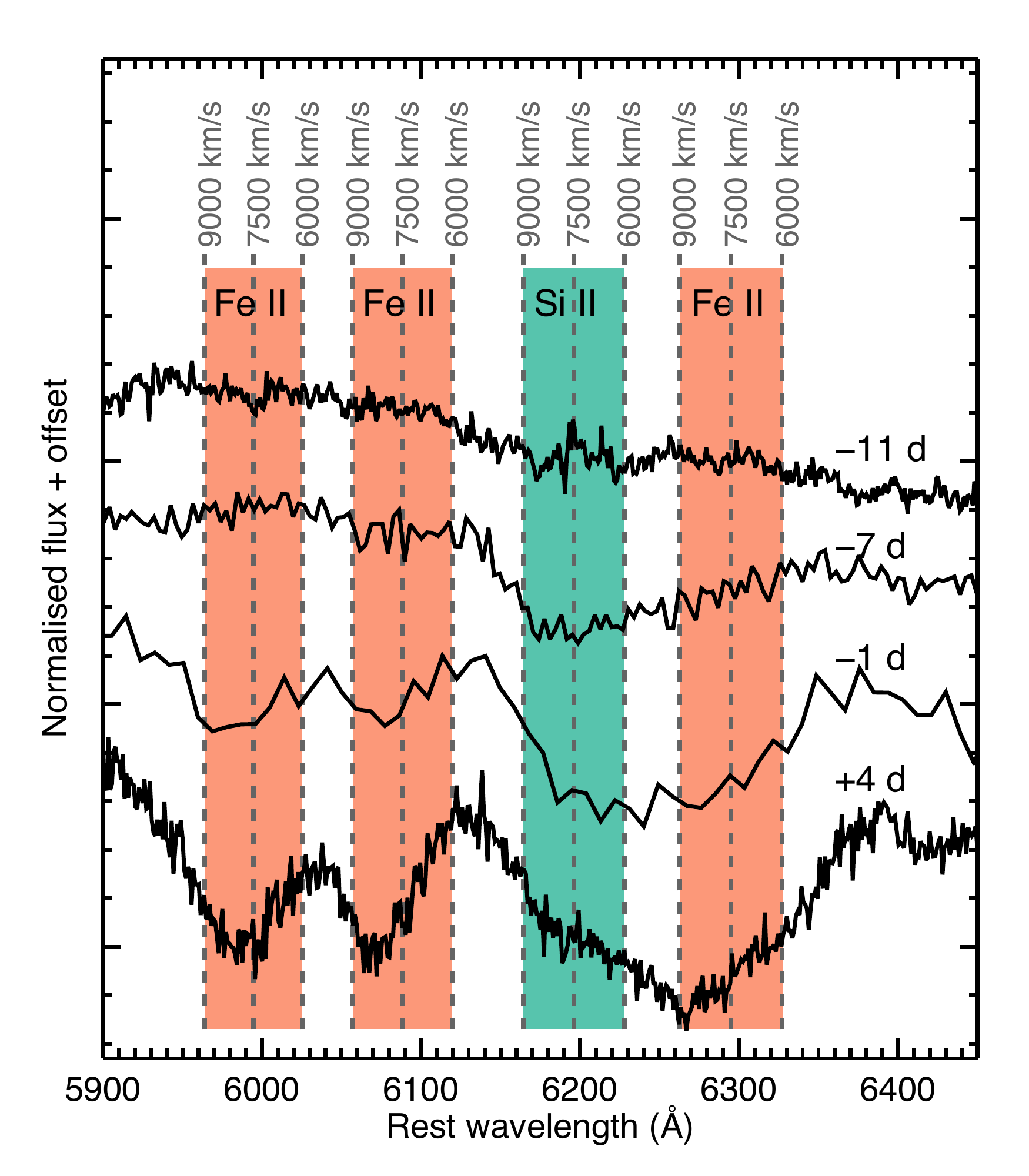}
    \caption{The \SiII\ and \Feii\ dominated region of the early ($-$11 to +4 d) spectra of SN 2020udy from $\sim$5900 -- 6450 \AA. Three \Feii\ features at 6148, 6244, 6456 \AA\ are shown with three different velocities of 6000, 7500 and 9500 \kms\ marked. The \SiII\ 6355 \AA\ feature in the same velocity range is also shown. The \Feii\ is seen to appear by $-$1 d with respect to maximum light. }
    \label{fig:velo_20udy}
\end{figure}

\subsection{Early-phase spectral comparison}
\label{sec:early_phase}
The optical spectra of SN 2020udy compared to those of three SNe Iax (SNe 2005hk, 2012Z, 2014hk) at three epochs, $-$11, $-7$, and 0 d relative to \textit{g}-band peak  are shown in Fig.~\ref{fig:opt_spec_comp} \citep{2007PASP..119..360P,Blo12,2013ApJ...767...57F,2015A&A...573A...2S,2015ApJ...806..191Y,2016MNRAS.459.1018T}. The earliest spectrum of SN 2020udy was obtained at $-$13 d with respect to \textit{g}-band peak (2 d after estimated t$_{\mathrm{fl}}$) and is shown in the left panel of Fig.~\ref{fig:opt_spec_comp}. 
This earliest spectrum is very blue and dominated by \Feiii\ absorption, which is also seen in the overluminous class of SNe Ia, 91T-like events \citep{1992ApJ...384L..15F,1992AJ....103.1632P}. The earliest spectrum of SN 2020udy shows only weak \SiII\ but this grows slightly stronger by the $-11$ d spectrum. The $-$11 d spectrum is very similar to those of SNe 2005hk and 2012Z in terms of the presence of absorption features of \Feiii\ and \SiII, as well as having similar line velocities. The $-$7 d spectrum of SN 2020udy (middle panel of Fig.~\ref{fig:opt_spec_comp}) appears most similar to SN 2005hk with more prominent \SiII\ absorption and more blended features than seen in SNe 2012Z and SN 2014ck. The presence of \Feiii\ is still visible. The maximum light spectra of SN 2020udy at $-1$ and $+$3 d with respect to \textit{g}-band peak (right panel of Fig.~\ref{fig:opt_spec_comp}) look very similar to those of SNe 2005hk and 2012Z, while the SN 2014ck spectra show significantly lower velocities.

\subsubsection{Velocity evolution of \SiII\ and \Feii}
The \SiII\ 6355 \AA\ and \Feii\ 6148, 6244, 6456 \AA\ region is shown in Fig.~\ref{fig:velo_20udy} for four spectral epochs for SN 2020udy. Velocities are marked in a typical range for high- and intermediate-luminosity SNe Iax of 6000 -- 9000 \kms\ for the \Feii\ features  at 6148, 6244, 6456 \AA\ and the \SiII\ 6355 \AA\ \citep{2007PASP..119..360P,Blo12,2013ApJ...767...57F,2015A&A...573A...2S,2015ApJ...806..191Y}. Potential contamination of the \SiII\ 6355 \AA\ from \Feii\ 6456 \AA\ absorption is likely from as early as $-$7  d  with respect to \textit{g}-band peak brightness, visible through the asymmetric profile shape. A clear contribution from \Feii\ 6456 \AA\ to the \SiII\ from $-$1 d is observed and is confirmed by the \Feii\ absorption features at 6148 and 6244 \AA\ with similar velocities.  A contribution of \Feii\ to the \SiII\ feature from as early as $-$9 d with respect to \textit{V}-band maximum was identified previously for the SN Iax, SN 2011ay \citep{Szalai2015}. This contribution from \Feii\ makes measurements of the \SiII\ velocities difficult but from Fig.~\ref{fig:velo_20udy}, we estimated expansion velocities of the \SiII\ feature of 7500$\pm$1000 \kms\ and similar \Feii\ velocities. These velocities are similar to those of the similar luminosity events, SNe 2005hk and 2012Z that had expansion velocities of $\sim$6500 and 7500 \kms, respectively. SN 2014ck that was fainter at peak and faster evolving had significantly lower line velocities at peak of $\sim$3000 \kms\ \citep{2016MNRAS.459.1018T}.

\subsubsection{\Cii\ feature detection?}
\begin{figure}
	\includegraphics[width=\columnwidth]{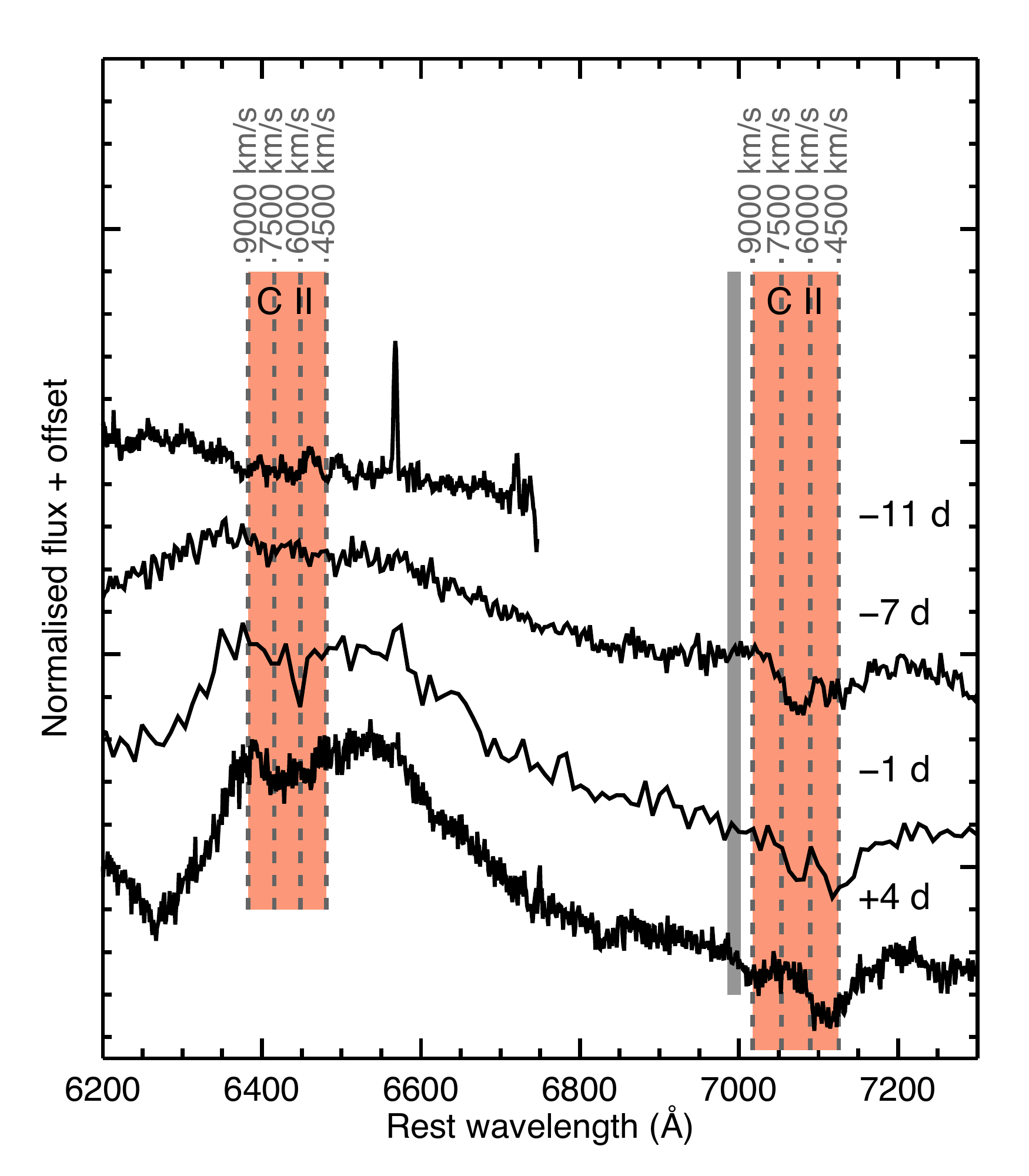}
    \caption{ The wavelength region covering the \Cii\ features at 6580 and 7234 \AA\ for four early spectra of SN 2020udy is shown. The typical velocities of 4500 -- 9000 \kms\ from the rest wavelengths of these lines are shown in the red regions with the vertical dashed gray lines showing the corresponding velocities. The solid gray dashed regions highlights a region with potential telluric features. }
    \label{fig:velo_cii}
\end{figure}

In SN 2020udy, we have identified a feature consistent with \Cii\ 6580 \AA\ at velocities of $\sim$7000$\pm$1000 \kms\ at phase of $-$11 to +4 d with respect to \textit{g}-band maximum (see Fig.~\ref{fig:velo_cii}). The wavelength region where \Cii\ 7234 \AA\ would be observed is more complex with likely at least two features contributing, making it more difficult to measure a clean velocity so we do not attempt to, but instead show the 4500 -- 9000 \kms\ velocity range in Fig.~\ref{fig:velo_cii}. The 7000$\pm$1000 \kms\ \Cii\ 6580 \AA\ line velocity is within the uncertainties of the \SiII\ measurement of 7500$\pm$1000 \kms\ but the contamination from other features makes a definitive conclusion on the presence of \Cii\ and its velocity difficult. The \Cii\ 6580 \AA\ feature appears to grow in strength with time, which is at odds with previous estimates that suggest it should be strongest pre-peak \citep{2013MNRAS.429.2287K}.

Lines associated with \Cii\ 6580 \AA, often \Cii\ 7234 \AA, and sometimes \Ci\ 9093, 9406 and 10693 \AA, have been suggested in many SNe Iax, with an estimate from \cite{2013ApJ...767...57F} of 80 -- 100 per cent of SNe Iax having \Cii\ pre-peak. However, there are puzzling aspects to the velocities obtained, e.g.~\cite{2015A&A...573A...2S} found that the \Ci\ line velocities in the near-infrared were $\sim$ 3000 \kms\ lower than the optical \Cii\ 6580 \AA\ feature. The velocities of \Cii\ are as low as half the \SiII\ velocity in the low-luminosity Type Iax SN 2019gsc \citep{2020MNRAS.496.1132T} and a little below the \SiII\ velocity in SNe 2012Z and 2014ck \citep{2015A&A...573A...2S,2016MNRAS.459.1018T}. These lower velocities of \Cii\ relative to \SiII\ have been suggested to be due to asymmetries in the distribution of \Cii\ in the ejecta, i.e.,~that it is clumped along of the line of sight \citep{2011ApJ...732...30P}. This is not inconsistent with the predictions of the N5def and its updated equivalent (`N5\_d2.6\_Z')  of \cite{lach22}, where their fig.~2 shows a somewhat asymmetric ejecta structure that is viewing-angle dependent. However, further multi-dimensional spectral modelling is required to investigate more thoroughly.

\subsection{Near-infrared photospheric spectral identifications}
\label{sec:nir_lineids}
\begin{figure}
	\includegraphics[width=\columnwidth]{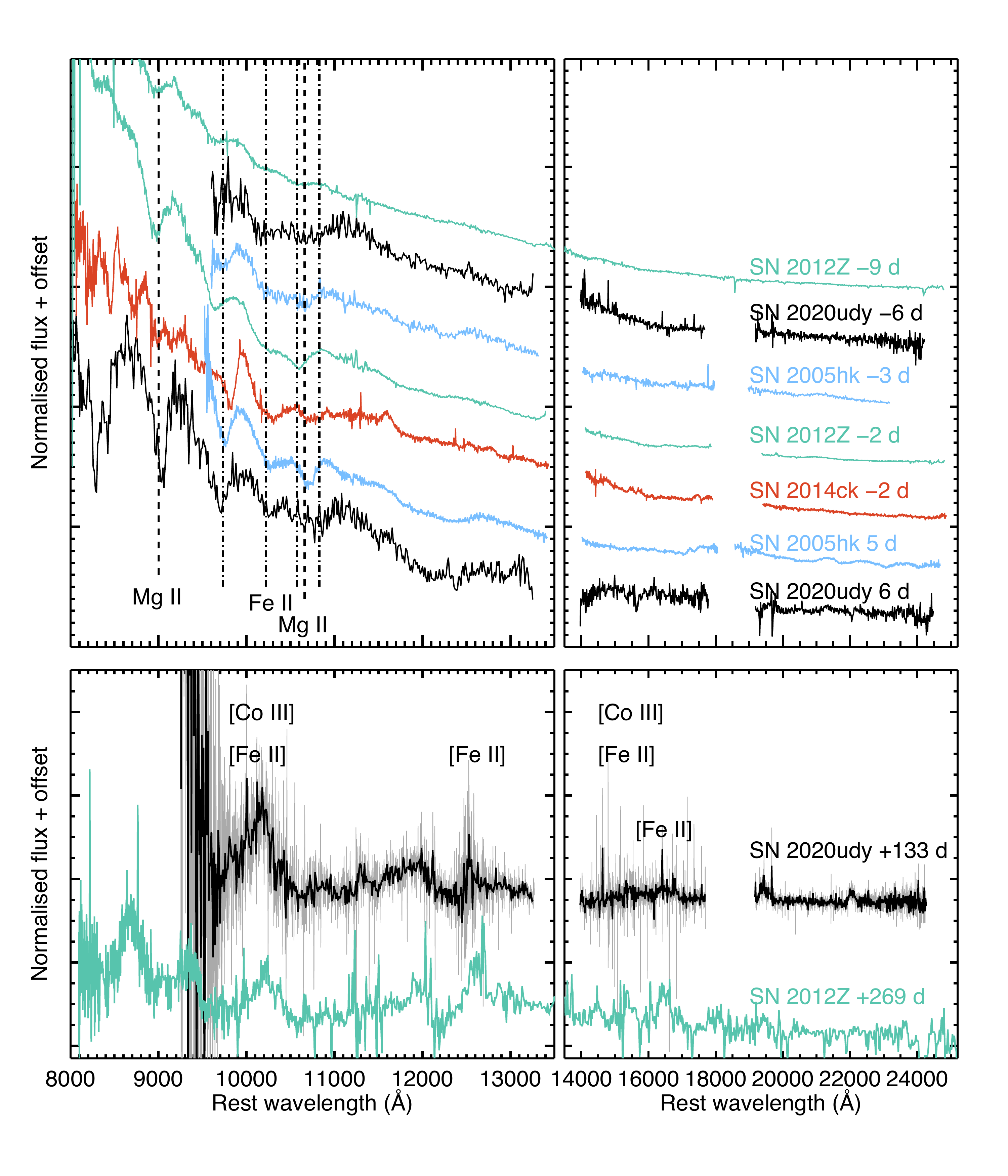}
    \caption{The top panel shows the $-$6 d and +6 d near-infrared spectra of SN 2020udy (black) are compared to those of three other 02cx-like SNe 2005hk, 2012Z, 2014ck, at similar phases. The positions of potential features of Mg II 9230 \AA\ (dashed) and Fe II 9998, 10500, 11126 \AA\ (dashed-dotted) are marked as vertical lines at a velocity of 8000 \kms. The bottom panel shows the near-infrared nebular phase spectrum of SN 2020udy at +133 d compared to that of SN 2012Z at +269 d. The black solid region shows the smoothed SN 2020udy while the grey region shows the original unsmoothed version.    }
    \label{fig:NIR_spec_comp}
\end{figure}

The two photospheric-phase near-infrared spectra of SN 2020udy, obtained at $-$6 and +6 d with respect to \textit{g}-band peak, are shown in Fig.~\ref{fig:NIR_spec_comp}, compared to similar phase spectra of SNe 2005hk \citep{2013MNRAS.429.2287K},  2012Z \citep{2015A&A...573A...2S}, and 2014ck \citep{2016MNRAS.459.1018T}.  Our line identifications were based on the previous lines identified in the papers on these objects, as well as from the \textsc{nist} atomic spectra database\footnote{https://www.nist.gov/pml/atomic-spectra-database}.  We have marked likely features associated with \Feii\ 9998, 10503, 10863, 11126 \AA\ and \Mgii\ 9231, 10927 \AA\ at velocities of 8000 \kms. The spectra of SN 2020udy are similar to those of other SNe Iax at similar epochs, with the most similarities between it and SN 2005hk. We searched for the presence of \Ci\ 9093, 9406 and 10693 \AA\ in these spectra of SN 2020udy, (only the +6 d spectrum covered $<9500$ \AA) but we did not identify \Ci\ absorption features in the spectra at velocities broadly consistent with the potential optical \Cii\ feature at 6580 \AA\ ($\sim$6000 -- 8000 \kms).

We have searched for the presence of \Hei\ 10830 and 20587 \AA\ absorption but find no evidence of it. There is also a lack of \Hei\ absorption features observed at optical wavelengths (\Hei\ 5876, 6678, 7065 \AA). \Hei\ has been identified in two SNe Iax, SNe 2004cs, 2007J \citep{2013ApJ...767...57F}, and tentatively detected at small abundance levels in SN 2010ae \citep{2019A&A...622A.102M}. However, whether SNe 2004cs and 2007J are true members of the SN Iax class is unclear \citep{2015ApJ...799...52W,2016MNRAS.461..433F}. In SN 2020udy, the region where a \Hei\ 10830 \AA\ feature would be expected is most similar to SN 2005hk, which also did not show \Hei\ absorption.

\begin{figure}
	\includegraphics[width=\columnwidth]{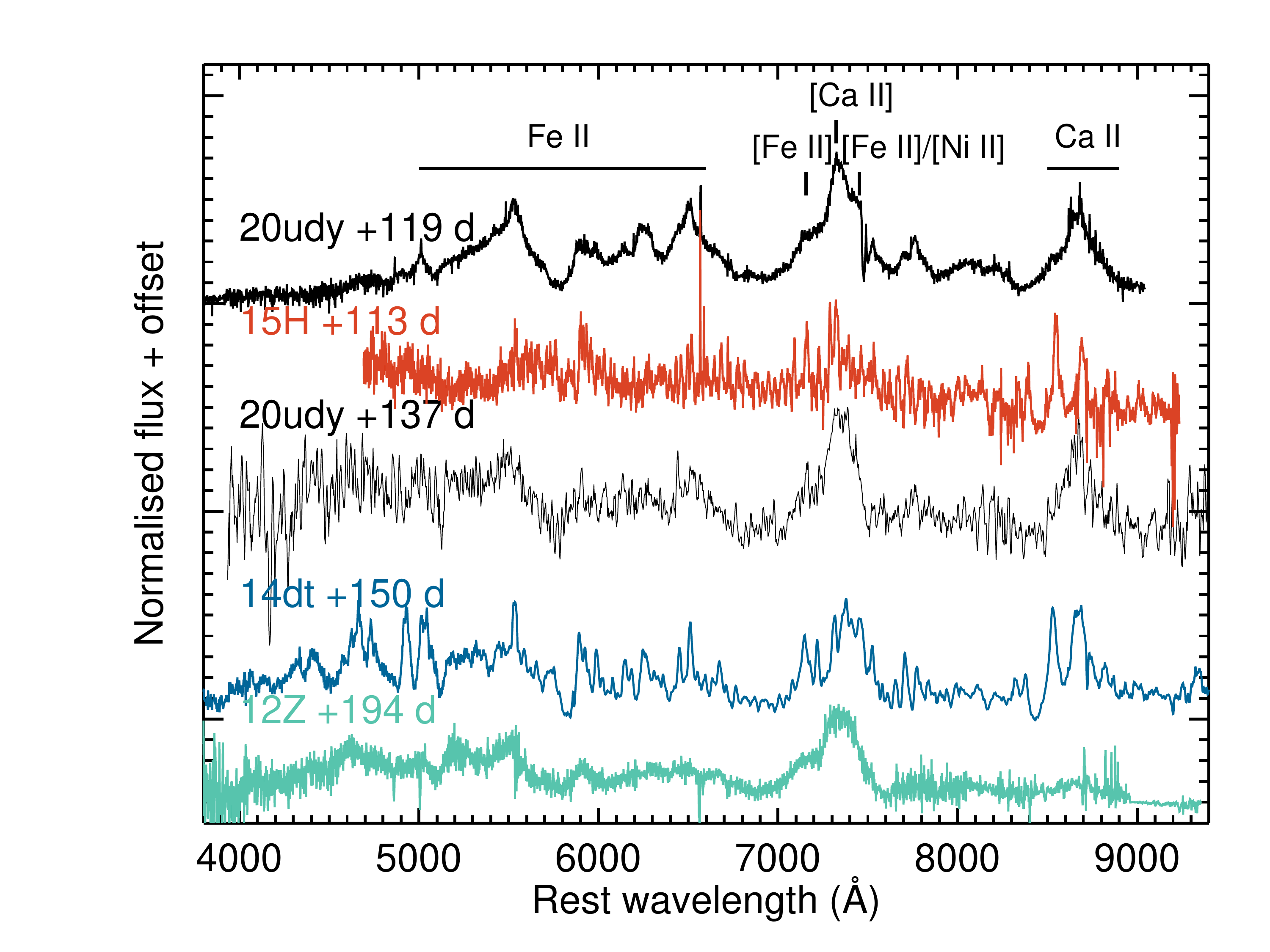}
    \caption{Comparison of the late-time optical (+119 d and +137 d) spectra of SN 2020udy to other SNe Iax with late-time spectra, SN 2014dt, SN 2015H, and SN 2012Z. SNe 2014dt and 2015H show low-velocity lines while those of SN 2012Z are more similar to those of SN 2020udy, with high expansion velocities that cause the lines to be more blended. Key line identifications are marked. The region from $\sim$5000 -- 6600 \AA\ is dominated by many overlapping lines of permitted \Feii.   }
    \label{fig:neb_spec}
\end{figure}

\subsection{Late-phase optical and near-infrared spectroscopy}
\label{latephase_spec}
The optical nebular phase (+119 d and +137 d) spectra of SN 2020udy are shown in Fig.~\ref{fig:neb_spec}, compared to those of three SNe Iax with spectra in the nebular phase (SNe 2014dt, 2015H, and 2012Z).  SN 2020udy is more similar to SN 2012Z \citep{2015A&A...573A...2S} than to SN 2014dt \citep{2023arXiv230203105C} and SN 2015H \citep{2016A&A...589A..89M} and has relatively high expansion velocities (seen through the widths of the emission features). SN 2012Z had a measured full width at half maximum of $\sim$9000 \kms\ and SN 2020udy has similar emission line widths.  The nebular phase spectra of SNe Iax are generally thought to be made of two distinct components \citep{2016MNRAS.461..433F}, i) standard nebular emission profiles coming from the ejecta and ii) a remaining photospheric component, which can be seen as blended permitted \Feii\ lines at $\sim$5000-6600 \AA. There is a wide variety in the nebular spectra of SN Iax with some events showing dominant broad emission lines, some showing only narrow features, and some with both. The late-time spectrum of SN 2020udy appears to be dominated by broad emission features expected to be coming from the SN ejecta, similar to SN 2012Z, while SN 2014dt and SN 2015H are dominated by narrower emission features. The detailed properties of the photospheric and forbidden emission lines of SN 2020udy will be analysed further in Camacho-Neves et al.~(in prep.).

The near-infrared nebular spectrum of SN 2020udy at +133 d from \textit{g}-band peak is shown in the bottom panel of Fig.~\ref{fig:NIR_spec_comp}, compared to a nebular spectrum of SN 2012Z at +269 d \citep{2015A&A...573A...2S}. The SN 2012Z spectrum is significantly later but prominent forbidden emission features of Fe-group elements are present in both objects. Due to the low S/N of the spectrum, we can not identify individual features but complexes typically associated with \CoiiiF\ and \FeiiF\ are seen at similar widths to those in SN 2012Z.

\section{Comparison to Chandrasekhar-mass deflagration models}
\label{sec:deflag_comp}

One of the most promising models to explain SNe Iax (or at least the intermediate to high-luminosity events) is the deflagration of a near Chandrasekhar-mass CO white dwarf \citep[e.g.][]{2004PASP..116..903B,2006AJ....132..189J}. Significant modelling efforts have been made to produce deflagration models for comparison to SNe Iax \citep{2013MNRAS.429.2287K,fink2014,lach22}. In Section \ref{def_lc}, we describe the light curve comparisons of SN 2020udy to these deflagration models and in Section \ref{def_spec}, we compare the spectra of SN 2020udy to the model spectra. 

\subsection{Light-curve comparisons to deflagration models}
\label{def_lc}

\begin{figure*}
	\includegraphics[width=13cm]{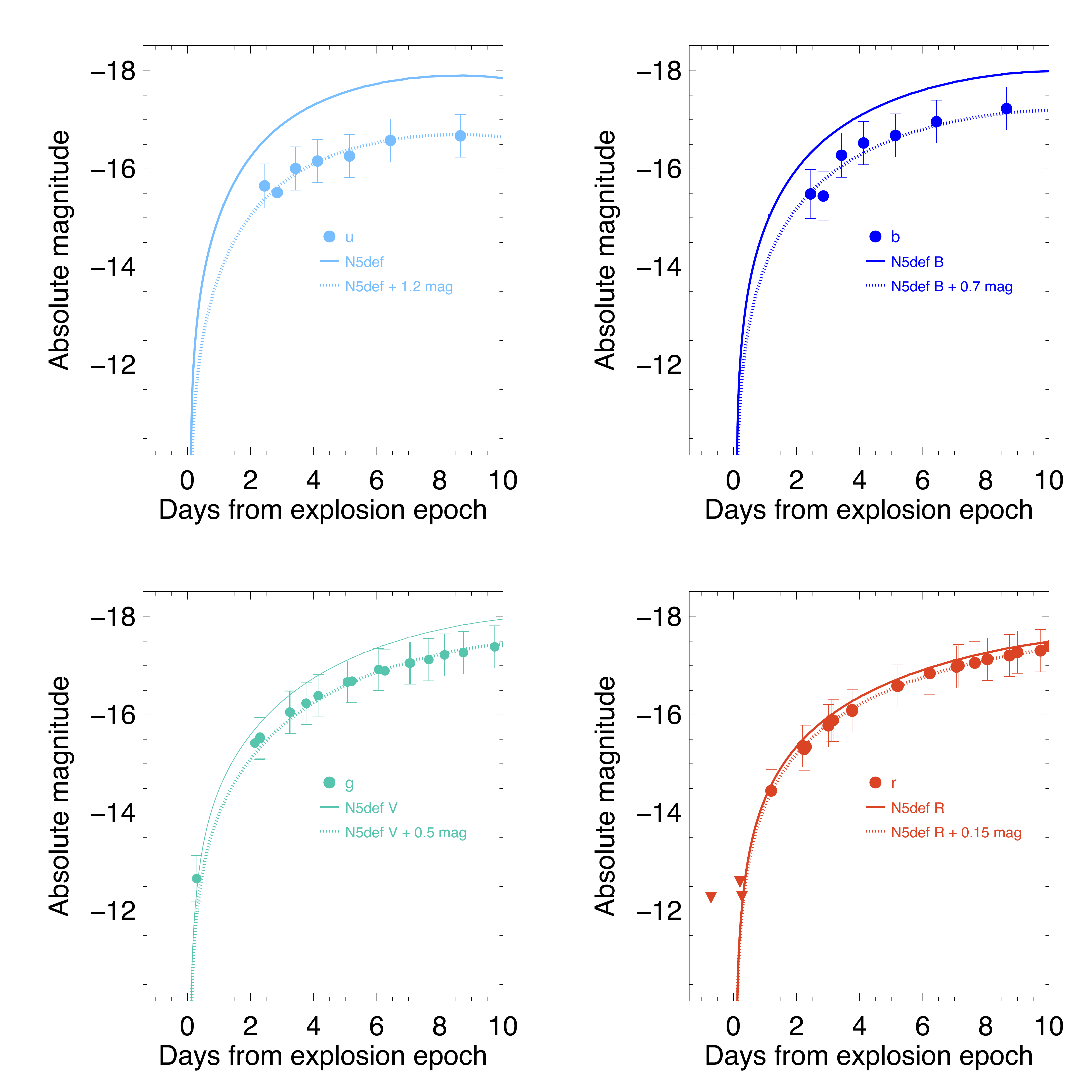}
    \caption{Absolute magnitude light curves of SN 2020udy in \textit{ubgr} bands compared to the early \textit{UBVR}-band N5def model \protect \citep{2013MNRAS.429.2287K} parameterised at early times by \protect \cite{2017MNRAS.472.2787N} as solid lines. The dotted lines show the N5def model offset, by the values listed in the legend, to best match the observed light curves in the  4 -- 10 d time range since explosion.  }
    \label{fig:early_evo}
\end{figure*}

In Fig.~\ref{fig:early_flux}, it can be seen that the early \textit{g}- and \textit{r}-band light curves of SN 2020udy are very similar to the N5def model. We measured the power-law index in the \textit{r} band, $\alpha_{\mathrm{r}}$, to be 1.29$\pm$0.07, which is smaller than the mean $\alpha_{\mathrm{r}}$ of  2.01$\pm$0.02 measured for normal SNe Ia \citep{2020ApJ...902...47M}. \cite{2017MNRAS.472.2787N} investigated the early light curve properties of a number of explosion models (delayed detonation, deflagration, double-detonation, merger) and found that increased mixing of the ejecta (e.g.~decreased stratification of elements) resulted in smaller $\alpha$ values. This is due to the radioactive material being present at high velocities in the ejecta. Therefore, light escapes earlier than for more confined \nick\ distributions, resulting in smaller $\alpha$ values. The near Chandrasekhar-mass deflagration model of a CO white dwarf, N5def, of \cite{2013MNRAS.429.2287K} is highly mixed due to the turbulent nature of the deflagration propagation. From a power-law fit to the model data from 2 -- 5 d post explosion, \cite{2017MNRAS.472.2787N} found $\alpha_{\mathrm{V}}$ to be 1.37 for the N5def deflagration model. Overall, we find that the shape of the early light curve rise (a proxy for the \nick\ distribution) of SN 2020udy is in excellent agreement with the N5def model light curves of \cite{2017MNRAS.472.2787N}. 

Figure \ref{fig:early_evo} shows the absolute magnitude light curves of SN 2020udy in the \textit{ubgr} bands compared to the absolute magnitude \textit{UBVR}-band light curves of the N5def model \citep{2013MNRAS.429.2287K} presented in \cite{2017MNRAS.472.2787N}. These models extend to 10 d after explosion. As discussed in Section \ref{sec:early_lc}, the model and observed filter responses are not exactly comparable but the \textit{g/V} and \textit{r/R} are in good agreement at differences of 0.07 and 0.02 mag, respectively. The first observed spectrum of SN 2020udy does not cover the wavelength region of the \textit{Swift} \textit{b} and Bessell \textit{B} filters. Therefore, to estimate the offset, we used the N5def model at +6.2 d that is in good agreement with the first spectrum of SN 2020udy (see Section \ref{def_spec}) and a difference of 0.01 mag was found for the \textit{b/B} band comparison. None of the observed or model spectra cover the wavelength range of the \textit{Swift} \textit{u} filter so the +6.2 d N5def model was extrapolated using a black body with the difference in the \textit{u/U} bands found to be $\sim$0.04 mag. These differences are smaller than the uncertainties on the estimate of the absolute magnitude of SN 2020udy and therefore, any disagreement between the model and observed photometry of SN 2020udy are unlikely to be dominated by differences in the filters responses. 

The N5def model \textit{UBVR} light curve shapes are very similar to the \textit{ubgr} band light curves of SN 2020udy (Fig.~\ref{fig:early_evo}), as was discussed for the early \textit{gr}-band only light curves in Section \ref{sec:early_lc}. The model light curves are seen to be brighter overall than SN 2020udy. However, these model light curves are the angle-averaged light curves and a range of magnitudes was identified in \cite{2013MNRAS.429.2287K} depending on the viewing angle, with the largest differences in the bluer bands. Observationally, the offset is also largest in the \textit{u}-band and decreases towards longer wavelength, with offsets of 1.2, 0.7, 0.5 and 0.15 mag in the \textit{ubgr} bands, respectively. These offsets are calculated by the best match of the model to the observed light curves in the 4 -- 10 d since explosion range -- earlier data was excluded because it may contain signatures of interaction (see Section \ref{sec:constraints}). These offset values are larger than the offsets suggested by the filter differences. We investigated if this offset could be corrected by including a host galaxy extinction component but no suitable value of A$_{V}$ was found that would not over- or under-correct some of the bands. To avoid over-correcting the \textit{r}-band, only an \textit{E(B--V)} value of $<$0.05 mag is allowed. Given the lack of narrow \NaiD\ absorption in the spectra and the offset location at 15 kpc from the galaxy core, we have not applied any host galaxy extinction correction. 

As seen in Fig.~\ref{fig:lc_comp}, the UV and optical light curves of SN 2020udy are very similar to SN 2005hk in terms of absolute brightness and evolution. \cite{2013MNRAS.429.2287K} showed a detailed comparison of the N5def model light curves to SN 2005hk and found that the peak brightness and colour at maximum were in reasonable agreement. Both the model and the observed objects (SNe 2005hk, 2020udy) lack a secondary maximum in their light curves, which is seen in ejecta with a high level of mixing of the Fe-group elements \citep{2006ApJ...649..939K,2013MNRAS.429.2287K}. However, there are discrepancies between the deflagration model and these SNe Iax; the N5def model was found to evolve too rapidly, both rising quicker and falling faster than SN 2005hk, and based on their similarity, also compared to SN 2020udy (see Fig.~\ref{fig:absmag_rise}). The rise time in the \textit{r} band for SN 2020udy is 21.3$\pm$0.5 d while the rise time for the N5def and the updated N5$\_$d2.6$\_$Z \citep{lach22} are 15.2 and 15.6 d, respectively. This is likely caused by a lower than required ejecta mass in the N5def model of  $\sim$0.4 \msun -- an increased ejecta mass would increase the gamma-ray trapping and slow down the evolution of the light curve \citep{2013MNRAS.429.2287K}. However, overall the shape and absolute magnitude predictions of the N5def model are in excellent agreement with the observations of the N5def model in the first 10 days post explosion. 

An additional component contributing to the light curve at late-times ($>$50-100 d) could come from the bound remnant in this scenario \citep{2017ApJ...834..180S} as was suggested for SN 2014dt \citep{kawabata_2018} to start to contribute at $\sim$70 d post peak. However, this is unlikely to be the origin of the moderate discrepancies between the N5def and SN 2020udy light curves at epochs of tens of days post maximum light. 

\subsection{Spectral comparison to deflagration models}
\label{def_spec}
\begin{figure}
	\includegraphics[width=\columnwidth]{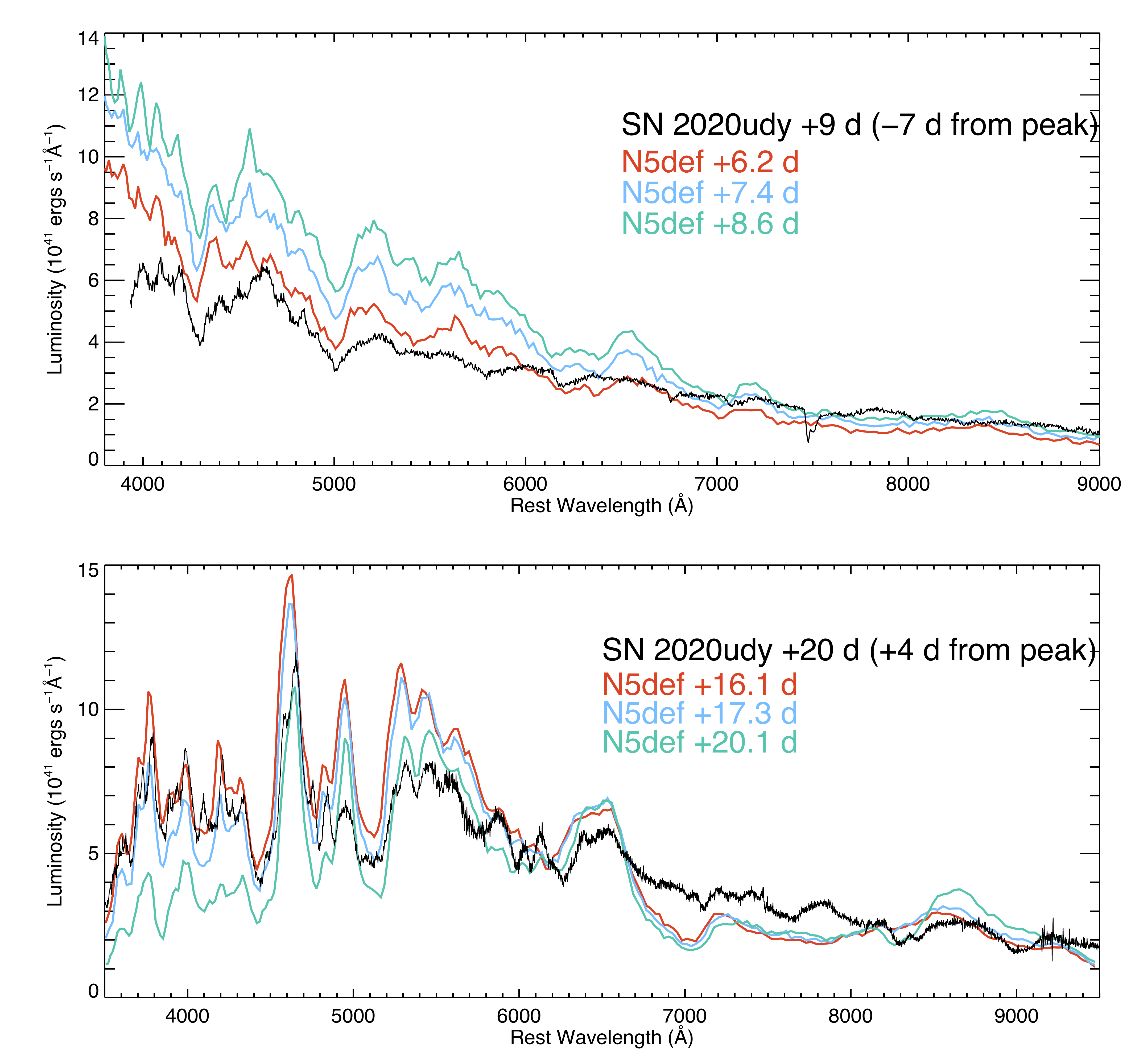}
    \caption{Spectral comparison between the $-$7 (+9 d from time of explosion $\sim t_{\rm fl}$) and +4 d (+20 d from time of explosion $\sim t_{\rm fl}$) spectra of SN 2020udy and the N5def deflagration model of \protect \cite{2013MNRAS.429.2287K} and \protect \cite{fink2014} at a number of epochs since explosion.  }
    \label{fig:def_comp}
\end{figure}

The earliest available N5def model spectrum is at +6.2 d after explosion. We have compared optical spectra of SN 2020udy at +9 and +20 d from t$_\mathrm{fl}$ (measured from power-law fits in Section \ref{sec:early_lc})  to the N5def spectral models at comparable epochs in Fig.~\ref{fig:def_comp} in luminosity space. The models are parameterised from time of explosion and we make the assumption that the time of explosion and t$_\mathrm{fl}$ occur at the same time since the deflagration models of \cite{2017MNRAS.472.2787N} have typical `dark phases' between explosion and first light of $<$0.2 d, due to the high degree of mixing of the ejecta and the prompt release of photons. Therefore, in the following we will refer to the phases as from the time of explosion, $t_\mathrm{exp}$. 

For the spectrum of SN 2020udy at +9 d from $t_\mathrm{exp}$ ($-$7 d with respect to \textit{g}-band peak), we have compared to N5def models at +6.2 (the earliest available), +7.4 and +8.6 d from $t_\mathrm{exp}$. The best match is seen for the +6.2 d spectrum in terms of luminosity, colour, presence of features, and line strengths. The later spectra are worse matches because they are brighter (due to the faster evolution of the model compared to the data) than SN 2020udy and they also show stronger line profiles, such as the \Feii\ 6456 \AA\ (see also Fig.~\ref{fig:velo_20udy}), than seen in the +9 d from $t_\mathrm{exp}$ spectrum of SN 2020udy. A similar phase mismatch is seen for the second epoch at +20 d from $t_\mathrm{exp}$ (+4 d with respect to \textit{g}-band peak) for SN 2020udy, where the best matching model in terms of overall luminosity is the +16.1 d model. In both cases, the observed spectrum is most consistent with models 3--4 d earlier than the expected phase. Based on the \textit{g--r} colour evolution of the N5def model shown in Fig.~\ref{fig:lc_comp}, the model evolves more quickly to the red, suggesting that the temperature of the model drops more quickly, allowing the \Feii\ absorption lines to become more prominent earlier than seen in the data. 

As discussed in \cite{2013MNRAS.429.2287K}, because the full white dwarf is not burnt in the N5def model, the kinetic energy of the explosion is lower than for other classes of explosion models. From \cite{fink2014}, the maximum velocity in the N5def model at which ejecta is present is 12000 \kms, which would correspond to the blue wing of our line profiles. Blue wings are hard to measure given the blending of line profiles as can be seen for the complex line profiles in Fig.~\ref{fig:velo_20udy}, as well as discussed in detail in \cite{2022MNRAS.509.3580M}. The minimum of the line profile will have a velocity lower than this for all the spectra and the highest velocities will be present in the earliest spectra. The spectral resolution of the models is too low to do detailed velocity comparisons. However, overall, when taking the contamination of \Feii\ into account, the \SiII\ velocity at maximum measured from the minimum of the absorption profile of 7500$\pm$1000 \kms\ is broadly consistent with the N5def model predictions of $<$12000 \kms. 

The nebular spectra of SN 2020udy contain both broad forbidden features of \CaiiF, \FeiiF, and \NiIIF, as well as features consistent with permitted \Feii\ as have been seen in similar luminosity SNe Iax, such as SNe 2002cx, 2005hk and 2012Z \citep{2006AJ....132..189J,2015A&A...573A...2S}. A possible explanation for these distinct features in the context of the deflagration models is that the broad features originate in the SN ejecta, while the permitted features come from the incompletely disrupted remnant \citep[e.g.~][]{2006AJ....132..189J,2008ApJ...680..580S,2016MNRAS.461..433F,maeda22_iax}.

\subsection{Constraints on a non-degenerate companion star}
\label{sec:constraints}
\begin{figure*}
	\includegraphics[width=16cm]{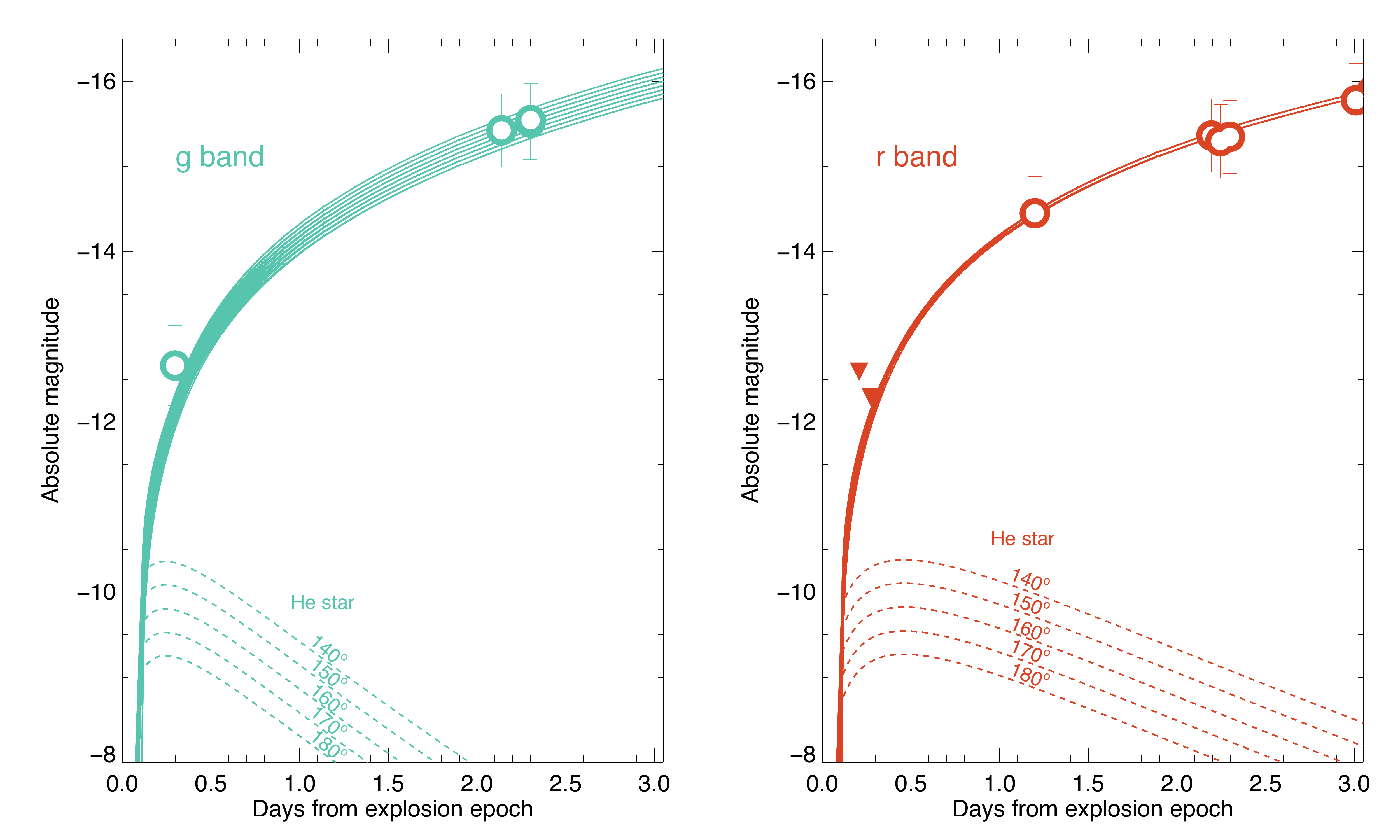}
    \caption{Absolute magnitude light curves in the \textit{g} (left panel) and \textit{r} bands (right panel) for SN 2020udy shown as data points in the first three days after explosion. These are compared to allowed models combining the N5def Chandrasekhar-mass deflagration model with the ejecta-companion interaction models of \protect \cite{kasen2010}. The allowed companion models within a 1-sigma range of the best-fitting one are shown. The interaction components are shown as dashed lines for different viewing angles, while the solid lines show the combined interaction and N5def models.   An offset to the N5def model was also allowed of up to 0.5 mag in the \textit{g}-band and 0.15 mag in the \textit{r}-band (based on the offsets in the fits to the later light curves discussed in Section \ref{def_lc}). The allowed models are a 1.2 \msun\ helium companion star based on \protect  \cite{2010ApJ...715...78P} at viewing angles of 140 -- 180$^{\circ}$, as well as the model with no companion interaction.  }
    \label{fig:def_companion}
\end{figure*}

Single-degenerate explosion models for SNe Ia, such as the deflagration of a near-Chandrasekhar mass white dwarf, suggest that there should be a non-degenerate companion star present. If this scenario is correct, then signatures of interaction between the SN Ia ejecta and this companion star may be identifiable as an additional flux contribution in the early light curves. 

The collision of SN Ia ejecta with main-sequence or red giant companion stars was investigated in \cite{kasen2010}.  The timescale and magnitude of interaction of the SN ejecta with a companion star are driven by the separation of the white dwarf and companion star - for mass transfer arising from Roche Lobe overflow, the separation is expected to be $\sim$3 times the radius of the companion star \citep{kasen2010}. In \cite{2014Natur.512...54M}, a bright blue source was identified in pre-explosion imaging of SN 2012Z and suggested to be due to a helium star, similar to the Milky Way helium nova V445 Puppis \citep{2008ApJ...684.1366K}. 

We have investigated the interaction of the SN ejecta with three companions: a 1.2 \msun\ helium-rich, a 2 \msun\ main-sequence, and a 6 \msun\ main-sequence star. There is a viewing angle dependence to the companion-interaction models, with the strongest signal seen when the SN is viewed from angles where the companion star is along the sight \citep[0$^\circ$; ][]{kasen2010}. We have used the viewing-angle dependence formulation of \cite{2015Natur.521..332O}, that has a much weaker, but still not zero, flux contribution when the explosion is viewed from angles directly opposite to the companion star (180$^\circ$). This parameterisation of the viewing-angle dependence is in agreement with the numerical simulations shown in fig.~2 of \cite{kasen2010}. The chosen helium-star radius of $\sim2$ $\times$ 10$^{10}$ cm is that of the maximum stellar radius used in the models of helium-star companion stars to SNe Ia in  \cite{2010ApJ...715...78P}. This value is also used for the analysis of interaction signatures in the early light curves of normal SN Ia, SN 2018aoz \citep{2023ApJ...946....7N}. We have assumed a constant opacity of $\kappa =$ 0.1 cm$^{2}$ g$^{-1}$ as in \cite{kasen2010}.  

We have combined the interaction models for each of the three companion stars with the early N5def light curve models of \cite{2017MNRAS.472.2787N}. A limitation of these combined models is that the N5def models are in the \textit{VR} and not in \textit{gr}. However, as discussed in Section \ref{def_lc}, the filter differences are small ($<$0.1 mag) in the optical bands studied. Therefore, we have directly combined the N5def \textit{V}-band light curve with the \textit{g}-band interaction light curves and compared them to the observed \textit{g}-band light curves of SN 2020udy. Similarly, the N5def \textit{R}-band model is combined with the \textit{r}-band interaction model for comparison with \textit{r}-band light curve of SN 2020udy. 

To account for the N5def model not being tuned to match the explosion properties of SN 2020udy, we have allowed an offset in both the \textit{g} and \textit{r} bands. This offset is allowed to be, in steps of 0.05 mag,  up to the offsets seen between the light curves of SN 2020udy and the N5def model (Fig.~\ref{fig:early_evo}), which corresponds to up to 0.5 mag in the \textit{g} band and up to 0.15 mag in the \textit{r} band. We have performed the comparison fits in absolute flux space so as to properly account for the early \textit{r} points that are $<$3$\sigma$ detections. Both bands are fit simultaneously. The interaction model viewing angles are in steps of 10$^{\circ}$ from 0$^{\circ}$ (with the companion star along the line of sight) to 180$^{\circ}$ (where the companion star is on the opposite of the SN). The best fitting models were estimated by evaluating the $\chi^2$ of each model fit to the \textit{g} and \textit{r}-band data up to three days post explosion. 

In Fig.~\ref{fig:def_companion}, we show the 1$\sigma$ range of allowed interaction models for SN 2020udy in the \textit{g} and \textit{r} bands in the first three days after estimated explosion. Overall, the best fitting model is one \textit{without} a contribution from any companion star with an offset for the N5def model of 0.35 mag in the \textit{V/g} and 0.15 mag in \textit{R/r}. The no companion models are allowed with offsets of 0.15--0.35 and 0.10 -- 0.15 mag in the \textit{V/g} and \textit{R/r} bands respectively. The 6 and 2 \msun\ main-sequence star models are not allowed at any viewing angle in the 1$\sigma$ range. For the helium star, viewing angles of 140--180$^{\circ}$ with respect to 0$^{\circ}$ for the companion star position are allowed  within the 1$\sigma$ range of the best model. The range of allowed offsets to the N5def models were 0.15--0.5 and 0.1--0.15 mag for the \textit{V/g} and \textit{R/r}-band models, respectively. 

The earliest \textit{gr}-band points at $<$0.5 d after explosion are most constraining for the models. If we exclude the data from this first epoch and start the fitting from 1 d after explosion then the number of possible fits increases significantly with all helium-star viewing angles allowed, 2 \msun\ main-sequence stars at viewing angles of 170 -- 180$^{\circ}$, and 6 \msun\ main-sequence stars at angles of 150 -- 180$^{\circ}$. 

\subsection{Constraints on asymmetries from polarimetry}
\cite{bulla_iax} modelled the spectropolarimetric signatures of the N5def model and the polarisation signature was found to be low overall. They also investigated the interaction of the SN ejecta in this scenario with a main-sequence companion star and found that there was an increased polarisation signature in the blue ($\lesssim5000$\,\AA) for viewing angles of 45$^{\circ}$ away from the companion star direction at maximum light, which was potentially seen in SN 2005hk \citep{bulla_iax}.  For SN 2020udy, we obtained imaging polarimetry at three epochs, at 0.1 and 7.0 d from maximum \textit{g}-band light in the \textit{BV} bands and at 30.6 d in the \textit{V} band. No signature of polarisation was detected in any of these observations at above the 3$\sigma$ level. Based on our comparison to the early N5def light curves (Section \ref{sec:constraints}), the light curves of SN 2020udy are consistent with either no interaction from a companion star or weak interaction with a helium star at viewing angles of 140 -- 180$^{\circ}$. For the N5def polarisation models, low polarisation ($P<0.1$ per cent) is expected at these viewing angles, in agreement with our non-detections.

\section{Discussion}
\label{sec:discussion}
We have presented an analysis of a comprehensive dataset of the nearby SN Iax, SN 2020udy, from the first detection at 7.2$\pm$0.4 hr after estimated first light until +150 d after maximum light. In the following sections, we put SN 2020udy in the context of other SNe Iax in terms of SN and host galaxies properties (Section \ref{sec:disc_comp}), discuss the promising model of the deflagration of a Chandrasekhar mass white dwarf that leaves behind a remnant (Section \ref{sec:disc_def}), as well as discuss the alternative model of a pulsational delayed detonation of Chandrasekhar mass white dwarf (Section \ref{sec:disc_pulso}).

\subsection{Comparison of SN 2020udy to other SNe Iax}
\label{sec:disc_comp}

The peak magnitude of SN 2020udy in the optical and UV, its rise and decay time post peak, and optical and UV colour evolution are most comparable to those of the well-studied SN Iax, SN 2005hk. We found rest-frame rise times for SN 2020udy of 15.6$\pm$0.7 and 21.3$\pm$0.5 d in the \textit{g} and \textit{r} bands, respectively, which sit in the `luminous' group of SNe Iax (Fig.~\ref{fig:absmag_rise}). The peak in the \textit{r} band is $\sim$6 d later than the peak in the \textit{g} band for SN 2020udy. For some fainter SNe Iax, this delay is shorter at just 1 -- 2 d. We have searched for a correlation between the absolute peak magnitude and this time delay in a similar sample to that shown in Fig.~\ref{fig:absmag_rise} but found no clear trend. However, these measurements can be hard to make to within a few days accuracy without high-cadence observations, e.g, see the typical uncertainties in Fig. \ref{fig:absmag_rise}. A longer delay between the peak of bluer and redder bands was seen for more luminous compared to less luminous deflagration models \citep{fink2014}. For the intermediate-luminosity deflagration models of \cite{lach22}, a delay of $\sim$3.5 -- 4.5 d from \textit{g}- to \textit{r}-band peak was found for the brighter models (e.g. N5def-like models), while for most of the fainter models this was closer to $\sim$2 d. 

The early light-curve rise of SN 2020udy can be fit with a power-law with indices of 1.38$\pm$0.11 and 1.29$\pm$0.07 in the \textit{g} and \textit{r} bands, respectively. A \textit{R}-band power-law index of 0.56$\pm$0.07 was found for the SN Iax, SN 2014ek \citep{2018MNRAS.478.4575L} , a \textit{r}-band power-law index of $\sim$1.3 was found for SN Iax, SN 2015H \citep{2016A&A...589A..89M}, and power-law indices in the \textit{gr} bands close to unity were found for the SN Iax, SN 2018cxk, which peaked at $-$17 mag in the \textit{R} band \citep{2020ApJ...902...47M}. These values are below the mean \textit{r}-band rise time of 2.01$\pm$0.02 of normal SNe Ia \citep{2020ApJ...902...47M}. This suggests a higher degree of mixing of the ejecta in SNe Iax \citep{2017MNRAS.472.2787N}, although the sample size of events with well-sampled light curves in the first days after explosion is small.

The earliest optical spectra of SN 2020udy from $-$13 d  with respect to \textit{g}-band maximum light ($\sim$2.3 d after t$_{\mathrm{fl}}$) are dominated by \Feiii\ and \SiII\ features, which quickly evolve to show \Feii\ absorption and potential \Cii\ -- the \Cii\ identification is uncertain due to likely contamination from other features. These are similar to the spectra of SNe 2005hk and 2012Z at similar epochs and the expansion velocities of the \SiII\ around peak in SN 2020udy of 7500$\pm$1000 \kms, are broadly consistent with those of SNe 2005hk and 2012Z of $\sim$6500 and 7500 \kms, respectively \citep{2007PASP..119..360P,2015A&A...573A...2S}.  We have not identified any feature consistent with \Hei\ 10830 \AA\ in the near-infrared spectra at $-$6 and +6 d with respect to \textit{g}-band maximum light. The late-time optical and near-infrared spectra at +119 and +133 d with respect to maximum light are most similar to those of SN 2012Z \citep{2015A&A...573A...2S}, with both displaying broad forbidden and permitted emission line profiles with widths of a few thousand \kms. 

The metallicity at the position of SN 2020udy has a value significantly below solar metallicity of 12 + log(O/H) of $<$ 8.30 (see Section \ref{sec:host}). This is lower than the values for SN 2012Z \citep{2015A&A...573A...2S}, PS1-12bwh \citep{2017A&A...601A..62M} and SN 2020sck \citep{2022ApJ...925..217D} that have similar or higher luminosities and have host galaxy metallicity measurements around the solar value. Some of the faint SNe Iax events, SNe 2008ha \citep{2009AJ....138..376F}, 2010ae \citep{2014A&A...561A.146S} and 2019gsc \citep{2020ApJ...892L..24S}, have been found to have low-metallicity environments (12 + log (O/H) $<$ 8.4) and it was suggested that there may be a preference for lower luminosity SNe Iax (fainter than $-$15.5 mag in \textit{r}/\textit{R}-band) to occur more frequently in low-metallicity environments \citep{2020ApJ...892L..24S}. However, there are exceptions of low-luminosity events occurring in more metal-rich environments, e.g., SN 2007J, SN 2010el \citep{2018MNRAS.473.1359L}, and SN 2020kyg  \citep{srivastav_iaxrates}. Overall, the environments of SNe Iax are consistent with on-going star formation suggesting a young stellar population  \citep{2018MNRAS.473.1359L} and a progenitor channel involving the explosion of near Chandrasekhar mass white dwarfs. SN 2020udy falls in the lower metallicity regime but is still consistent with the median SNe Iax metallicity of $\sim8.4$ \citep{2018MNRAS.473.1359L}. 

\subsection{The deflagration of a near Chandrasekhar mass white dwarf}
\label{sec:disc_def}
In Section \ref{sec:deflag_comp}, we performed a detailed comparison of the observations of SN 2020udy to the light curve and spectral models of the N5def deflagration model of a near Chandrasekhar mass white dwarf \citep{2013MNRAS.429.2287K,2017MNRAS.472.2787N}. We found good agreement in both the early and peak magnitudes of SN 2020udy compared to the N5def light curves, as well as excellent agreement in the shape of the early (0--6 d from time of first light) light curve (Fig.~\ref{fig:early_flux}) when normalised to peak. In Fig.~\ref{fig:early_evo}, we showed the light curves for 0--10 d after time of first light in absolute magnitude space (this was the time range covered by the early models). An offset was applied to provide better agreement between the N5def models and SN 2020udy, with offset decreasing from \textit{u} to \textit{r} bands from 1.2 to 0.15 mag, respectively. This offset could not be explained by an additional host galaxy extinction component and was also larger than any expected filter differences. The evolution of the N5def models is faster than that of SN 2020udy (and other similar events such as SN 2005hk) with rise times of 5 -- 6 d shorter than for SN 2020udy. This implies insufficient trapping of radiation in the ejecta and was suggested by \cite{2013MNRAS.429.2287K} to be caused by a lower than required ejecta mass in the model. Differences in radiative transfer codes have recently been investigated in detail and it is plausible that these offset and differences may be reduced (or increased) by the use of different codes in the future \citep{Blondin_compareRT}.

The potential contribution of the collision of the SN ejecta with a non-degenerate companion star to the early light curves was investigated for a helium star based on \cite{2010ApJ...715...78P}, as well as 2 and 6 \msun\ companion stars  \citep{kasen2010}. The viewing-angle dependence of the companion interaction was parameterised as in \cite{2015Natur.521..332O}. The best-fitting combination of the N5def and companion star models were estimated simultaneously in absolute flux space  for the \textit{g} and \textit{r} bands. The best-fitting model is that without any contribution from companion-star interaction but models of a helium-companion star, with viewing angles of 140 -- 180$^{\circ}$ are allowed within the 1$\sigma$ range of the lowest $\chi^2$ model (Fig.~\ref{fig:def_companion}). Interaction with a 2 and 6 \msun\ main-sequence companion models are excluded along with helium companion models from 0 -- 130$^{\circ}$. A major caveat in this companion analysis is that the models of \cite{kasen2010} assume a Chandrasekhar mass explosion, while the N5def model has an ejecta mass of $\sim$0.4 \msun\ leaving a remnant of $\sim$1 \msun. This would result in a less energetic explosion interacting with the companion material and likely means that more helium-rich and main-sequence companion models could be allowed. More extensive modelling of this parameter space would allow stricter constraints to be placed for SN 2020udy but is beyond the scope of this paper. 

A potential helium star companion was identified by \cite{2014Natur.512...54M} in pre-explosion imaging of SN 2012Z and pre-explosion detection limits are consistent with this companion type \citep{2010AJ....140.1321F,2015ApJ...798L..37F}. Very late-time ($\sim$1400 d after explosion) images of SN 2012Z still contain a source that is brighter than the pre-explosion limits, suggesting a contribution from a companion star, interaction with circumstellar material or the contribution from a remnant star \citep{mccully22}.  No helium was identified in the near-infrared spectrum of SN 2020udy in common with some other more luminous SNe Iax (SNe 2005hk, SN 2012Z, SN 2015H) with near-infrared data, with helium abundance limits of $<10^{-3}$ \msun\ \citep{2019A&A...622A.102M}. 

\subsection{The pulsational delayed-detonation model of Chandrasekhar-mass white dwarf}
\label{sec:disc_pulso}

The pulsational delayed detonation model of a Chandrasekhar-mass white dwarf \citep[PDDEL;][]{1991A&A...245L..25K,1993A&A...270..223K,1995ApJ...444..831H, 2014MNRAS.441..532D} has also been suggested to be a potential explosion mechanism for SNe Iax, such as SN 2012Z \citep{2015A&A...573A...2S}. This model begins with an initial sub-sonic deflagration phase that does not unbind the white dwarf and results in the fall back of material that triggers a subsequent detonation. Key observable predictions of a relevant low \nick\ (0.253 \msun) mass version of this model  \citep[e.g.~PDDEL12 of ][]{2014MNRAS.441..532D} is early blue colours, strong \Cii\ signatures in the early spectra, relatively stratified ejecta produced in a detonation phase, and slow velocity evolution of 14000 to 12000 \kms\ from explosion to maximum \textit{g}-band peak.   A companion star to the Chandrasekhar mass white dwarf would also be present in this scenario. The observations of SN 2020udy are at least qualitatively in contrast with these models in terms of the colours  \cite[see discussion in][]{miller_16fnm}  and the lack of stratification seen in the ejecta \citep{2022MNRAS.509.3580M}. However, without available hydrodynamic simulations of the early light curves from a PDDEL model a direct comparison with the observations is not possible. 

\section{Conclusions}
\label{sec:conclusions}

We have presented a detailed analysis of the UV and optical light curves of the luminous SN Iax, SN 2020udy, with the first light curve detection at just 0.3$\pm$0.1 (7.2$\pm$0.4 hr) after estimated first light from a power-law fit to the early optical light curves. Our first optical spectrum is at $\sim$2 d after first light ($-13$ d with respect to \textit{g}-band peak) and optical/near-infrared spectral coverage extended until $\sim$130 d after peak. Broad-band optical linear polarisation measurements at and after peak were also obtained. Our main results are:

\begin{enumerate}
    \item The optical and UV light curves of SN 2020udy are similar in absolute magnitude and evolution to the class of `luminous' SNe Iax and are closest to the properties of SN 2005hk.
    \item Its early \textit{gr}-band light curves are best fit with power-law indices  significantly smaller ($\alpha_{\mathrm{g}}$ = 1.38$\pm$0.11, $\alpha_{\mathrm{r}}$ = 1.29$\pm$0.07) than the mean for `normal' SNe Ia of $\alpha_{\mathrm{r}}$ = 2.01$\pm$0.02 \citep{2020ApJ...902...47M}, suggestive of more extensive \nick\ mixing \citep{2017MNRAS.472.2787N}.
    \item The early (0--10 d from explosion) SN 2020udy light curves are well described by the N5def deflagration model \citep{2013MNRAS.429.2287K} but a flux/magnitude offset is required.
    \item As found in other luminous SN Iax studies, the light curves, colour curves, and spectra of SN 2020udy evolve more slowly than those of the N5def model.
    \item A contribution to its early light curves from a non-degenerate (6 \msun, 2 \msun\ main-sequence or a 1.2 \msun\ helium star) is ruled out for all but viewing angles of 140 - 180$^{\circ}$ for the helium companion star. 
    \item No significant broad-band linear polarisation signature was identified at three epochs  in the \textit{BV} bands spanning from \textit{g}-band peak to +30 d.
    \item The optical and near-infrared spectra of SN 2020udy are similar to those of SNe 2012Z and 2005hk, with \SiII\ velocities around peak of 7500$\pm$1000 \kms.
    \item A potential \Cii\ 6580 \AA\ is seen but it appears to grow stronger with time, suggesting it may not be due to \Cii.
    \item The late-time optical and near-infrared spectra of SN 2020udy show relatively broad forbidden and permitted features similar to those of SN 2012Z at similar epochs.  
    \item The galaxy environment of SN 2020udy is low metallicity ($<$0.41 \zsun). 
\end{enumerate}

SN 2020udy is the SN Iax with the earliest detection after estimated time of first light ($\sim$7 hr) and so has the strictest constraints on a potential companion star interaction. Its light curve and spectral properties are consistent with the failed deflagration of a Chandrasekhar-mass white dwarf with a helium companion star. However, no isolated light curve feature due to this interaction is detected and the light curves are most consistent with no companion interaction. This suggests that at least one of the following is true: i) we were unlucky with our viewing angle of the ejecta-helium star interaction and viewed it at the angle least favourable for a prominent signature, ii) the models of \cite{kasen2010} overpredict the interaction signature in this scenario of a low-energy deflagration combined with a helium companion star, or iii) the failed deflagration of a near-Chandrasekhar mass white dwarf is not the explosion model for SNe Iax. We believe this last possibility to be the least likely given the extensive and robust evidence in the literature to explain this sub-class of SNe Ia with this model. However, further detailed observations of SNe Iax within hours of explosion, as well as modelling of the interaction of a sub-sonic explosion with helium companion stars with a range of properties, are required to place more stringent constraints on this scenario.

\section*{Acknowledgements}
KM and GD acknowledge support from EU H2020 ERC grant no. 758638. MRM acknowledges a Warwick Astrophysics prize post-doctoral fellowship made possible thanks to a generous philanthropic donation. GL and MP are supported by a research grant (19054) from VILLUM FONDEN. KD was supported by NASA through the NASA Hubble Fellowship grant \#HST-HF2-51477.001 awarded by the Space Telescope Science Institute, which is operated by the Association of Universities for Research in Astronomy, Inc., for NASA, under contract NAS5-26555. AGY's research is supported by the EU via ERC grant No. 725161, the ISF GW excellence center, an IMOS space infrastructure grant and a GIF grants, as well as the André Deloro Institute for Advanced Research in Space and Optics, The Helen Kimmel Center for Planetary Science, the Schwartz/Reisman Collaborative Science Program and the Norman E Alexander Family M Foundation ULTRASAT Data Center Fund, Minerva and Yeda-Sela;  AGY is the incumbent of The Arlyn Imberman Professorial Chair. YY is supported by a Bengier-Winslow-Robertson Fellowship. MMK acknowledges generous support from the David and Lucille Packard Foundation.  

Based on observations obtained with the Samuel Oschin Telescope 48-inch and the 60-inch Telescope at the Palomar Observatory as part of the Zwicky Transient Facility project. ZTF is supported by the National Science Foundation under Grants No. AST-1440341 and AST-2034437 and a collaboration including current partners Caltech, IPAC, the Weizmann Institute of Science, the Oskar Klein Center at Stockholm University, the University of Maryland, Deutsches Elektronen-Synchrotron and Humboldt University, the TANGO Consortium of Taiwan, the University of Wisconsin at Milwaukee, Trinity College Dublin, Lawrence Livermore National Laboratories, IN2P3, University of Warwick, Ruhr University Bochum, Northwestern University and former partners the University of Washington, Los Alamos National Laboratories, and Lawrence Berkeley National Laboratories. Operations are conducted by COO, IPAC, and UW. SED Machine is based upon work supported by the National Science Foundation under Grant No. 1106171. This work was supported by the GROWTH project \citep{Kasliwal2019} funded by the National Science Foundation under Grant No 1545949. 

 The Liverpool Telescope is operated on the island of La Palma by Liverpool John Moores University in the Spanish Observatorio del Roque de los Muchachos of the Instituto de Astrofisica de Canarias with financial support from the UK Science and Technology Facilities Council. Based on observations made with the Nordic Optical Telescope, owned in collaboration by the University of Turku and Aarhus University, and operated jointly by Aarhus University, the University of Turku and the University of Oslo, representing Denmark, Finland and Norway, the University of Iceland and Stockholm University at the Observatorio del Roque de los Muchachos, La Palma, Spain, of the Instituto de Astrofisica de Canarias. W.\ M.\ Keck Observatory and MMT Observatory access was supported by Northwestern University and the Center for Interdisciplinary Exploration and Research in Astrophysics (CIERA). Some of the data presented here were obtained with the Visiting Astronomer facility at the Infrared Telescope Facility, which is operated by the University of Hawaii under contract 80HQTR19D0030 with the National Aeronautics and Space Administration. Some of the data presented herein were obtained at the W. M. Keck Observatory, which is operated as a scientific partnership among the California Institute of Technology, the University of California and the National Aeronautics and Space Administration. The Observatory was made possible by the generous financial support of the W. M. Keck Foundation. The authors wish to recognize and acknowledge the very significant cultural role and reverence that the summit of Maunakea has always had within the indigenous Hawaiian community.  We are most fortunate to have the opportunity to conduct observations from this mountain. This work made use of the Heidelberg Supernova Model Archive (HESMA), https://hesma.h-its.org.

The Pan-STARRS1 Surveys (PS1) and the PS1 public science archive have been made possible through contributions by the Institute for Astronomy, the University of Hawaii, the Pan-STARRS Project Office, the Max-Planck Society and its participating institutes, the Max Planck Institute for Astronomy, Heidelberg and the Max Planck Institute for Extraterrestrial Physics, Garching, The Johns Hopkins University, Durham University, the University of Edinburgh, the Queen's University Belfast, the Harvard-Smithsonian Center for Astrophysics, the Las Cumbres Observatory Global Telescope Network Incorporated, the National Central University of Taiwan, the Space Telescope Science Institute, the National Aeronautics and Space Administration under Grant No. NNX08AR22G issued through the Planetary Science Division of the NASA Science Mission Directorate, the National Science Foundation Grant No. AST-1238877, the University of Maryland, Eotvos Lorand University (ELTE), the Los Alamos National Laboratory, and the Gordon and Betty Moore Foundation.

\section*{Data Availability}

All observations will be made public via WISeREP (https://www.wiserep.org/).
 



\bibliographystyle{mnras}
\bibliography{SN2020udy} 



 \appendix

\section{Additional spectroscopy}
\begin{table}
	\centering
	\caption{Additional spectra of SN 2020udy that are not used in the analysis in this paper due to phase overlap with other higher S/N data. They are available on WISeREP.}
	\label{tab:spec_obs_add}
	\begin{tabular}{lccc} 
		\hline
		Night of obs. & MJD$^a$ & Phase (d)$^b$ & Telescope+instrument \\
		\hline
        20200926 &59118.2& $-$12.6&P60+SEDM \\
        20200928 & 59120.1 & $-$10.7 & LT+SPRAT \\
        20200928 &59120.4 &$-$10.4&P60+SEDM \\
        20201027 &59149.2 & +17.9& P60+SEDM \\
        20201028 &59150.1&+18.8&P60+SEDM \\
        20201030 &59152.2& +18.9&P60+SEDM \\
        20201108 &59161.2 &+29.7& LT+SPRAT \\
        20201116 &59169.0 &+37.4& LT+SPRAT \\  
	  \hline
	\end{tabular}
 \begin{flushleft}
$^a$MJD = Modified Julian date.\\
$^b$Rest frame phase relative to $g$-band maximum light of 59131.0$\pm$0.7.\\
 \end{flushleft}
\end{table}


\bsp	
\label{lastpage}
\end{document}